\documentclass[reqno, 11pt]{amsart}

\usepackage{hyperref, cleveref, amsmath, amsthm, bm}
\usepackage[authoryear,comma]{natbib}
\usepackage{amsaddr}
\usepackage{graphicx}
\usepackage{enumitem}
\usepackage{accents}
\usepackage{algpseudocode, algorithm}
\usepackage{subcaption}

\usepackage[margin=1.2in]{geometry}
\usepackage{enumitem}

\newtheorem{theorem}{Theorem}
\newtheorem{proposition}{Proposition}

\newtheorem{assumption}{Assumption}
\crefname{assumption}{Assumption}{Assumptions}
\crefname{equation}{equation}{equations}

\theoremstyle{definition}

\newtheoremstyle{propertystyle}
{3pt} 
{3pt} 
{\it} 
{} 
{\bfseries} 
{.} 
{.5em} 
{} 
\theoremstyle{propertystyle}

\theoremstyle{remark}

\newcommand\independent{\protect\mathpalette{\protect\independenT}{\perp}}
\def\independenT#1#2{\mathrel{\rlap{$#1#2$}\mkern2mu{#1#2}}}

\newlength{\dhatheight}

\begin{document}

\title[Tailored Loss Functions]{Covariate Balancing Propensity Score by
Tailored Loss Functions}
\author{Qingyuan Zhao}
\address{Department of Statistics, Wharton School, University of Pennsylvania}
\email{qyzhao@wharton.upenn.edu}
\date{\today}
\thanks{The author thanks Trevor Hastie, Hera Y.\ He and Dylan Small
  for valuable comments.}

\maketitle

\begin{abstract}
  In observational studies, propensity scores are commonly estimated by
maximum likelihood but may fail to balance high-dimensional
pre-treatment covariates even after specification search. We introduce a
general framework that unifies and generalizes several recent
proposals to improve covariate balance when designing an observational
study. Instead of the likelihood function, we propose to optimize special
loss functions---covariate balancing scoring rules
(CBSR)---to estimate the propensity score. A CBSR is uniquely determined by the
link function in the GLM and the estimand (a weighted
average treatment effect). We show CBSR does not lose asymptotic
efficiency to the Bernoulli likelihood in estimating the weighted
average treatment effect compared, but CBSR is much more robust in
finite sample. Borrowing tools developed in statistical
learning, we propose practical strategies to balance covariate
functions in rich function classes. This is useful to estimate the
maximum bias of the inverse probability weighting (IPW) estimators and
construct honest confidence interval in finite sample. Lastly, we provide
several numerical examples to demonstrate the trade-off of bias and
variance in the IPW-type estimators and the trade-off in balancing
different function classes of the covariates.
\end{abstract}

\section{Introduction}
\label{sec:introduction}

To obtain causal relations from observational data, one crucial
obstacle is that some pre-treatment
covariates are not balanced between the treatment groups. Exact
matching, inexact matching and subclassification on raw covariates
were first used by pioneers like
\citet{cochran1953matching,cochran1968effectiveness} and
\citet{rubin1973matching}. Later in the seminal work of
\citet{rosenbaum1983}, the propensity score, defined as the
conditional probability of receiving treatment given the covariates,
was established as a fundamental tool to
adjust for imbalance in more than just a few covariates. Over the next
three decades, numerous methods based on the propensity score have
been proposed, most notably propensity score matching
  \citep[e.g.][]{rosenbaum1985constructing,abadie2006large},
  propensity score subclassification \citep[e.g.][]{rosenbaum1984}, and inverse
  probability weighting \citep[e.g.][]{Robins1994,Hirano2001}; see
  \citet{imbens2004nonparametric,Lunceford2004,caliendo2008some,stuart2010matching}
  for some comprehensive reviews.

With the rapidly increasing ability to collect high-dimensional
covariates in the ``big data'' era (for example large number of covariates
collected in health care claims data), propensity-score based methods
often fail to produce satisfactory covariate balance
\citep{Imai2008}. In the meantime, numerical examples in
\citet{smith2005does,Kang2007} have demonstrated that the average
treatment effect estimates can be highly sensitive to the working
propensity score model. Conventionally, these two issues are handled
by a specification search---the estimated propensity score is applicable only
if it balances covariates well. A simple strategy is to gradually
increase the model complexity by forward stepwise regression
\citep[Section 13.3--13.4]{Imbens2015}, but as a numerical example
below indicates, this has no guarantee to achieve sufficient
covariate balance eventually.

More recently, several new methods were proposed to directly improve
covariate balance in the design of an observational study, either by
modifying the propensity score model
\citep{graham2012inverse,Imai2014} or by directly constructing
sample weights for the observations
\citep{Hainmueller2011,hazlett2013balancing,zubizarreta2015stable,chan2015,kallus2016generalized}. These
methods have been shown to work very well empirically (particularly in
finite sample) and some asymptotic justifications were subsequently
provided \citep[e.g.][]{zhao2015,fan2016improving}.

In this paper, we will introduce a general framework that unifies and
generalizes these proposals. The solution provided
here is conceptually simple: in order to improve covariate balance of
a propensity score model, one just needs to minimize, instead of the
most widely used negative Bernoulli likelihood, a special loss function
tailored to the estimand.

\subsection{A toy example}
\label{sec:toy-example}

To demonstrate the simplicity and effectiveness of the tailored loss
function approach, we use the prominent simulation example of
\citet{Kang2007}. In this example, for each unit $i=1,\dots,n = 200$,
suppose that $(Z_{i1},Z_{i2},Z_{i3},Z_{i4})^T$ is independently
distributed as $\mathrm{N}(0,I_4)$ and the true propensity scores are $p_i =
P(T_i=1|Z_i) = \mathrm{expit}(-Z_{i1}+0.5Z_{i2}-0.25Z_{i3}-0.1Z_{i4})$
where $T_i \in \{0,1\}$ is the treatment label. However, the
observed covariates are nonlinear transformations of $Z$: $X_{i1} =
\exp(Z_{i1}/2)$, $X_{i2} = Z_{i2}/(1+\exp(Z_{i1})) + 10$, $X_{i3} =
(Z_{i1}Z_{i3}/25+0.6)^3$, $X_{i4}=(Z_{i2}+Z_{i4}+20)^2$. To model the
propensity score, we use a logistic model with some or all of
$\{X_{1},X_{2},X_{3},X_{4},X_{1}^2,X_{2}^2,X_{3}^2,X_{4}^2\}$ as
regressors. Using forward stepwise regression, two series of models
are fitted using the Bernoulli likelihood and the loss
function tailored for estimating the average treatment effect (ATE,
see \Cref{sec:covar-balanc-scor} for more detail). Inverse probability weights (IPW)
are obtained from each fitted model and standardized differences of
the regressors are used to measure covariate imbalance
\citep{rosenbaum1985constructing}.

\begin{figure}[t]
  \centering
  \includegraphics[width = \textwidth]{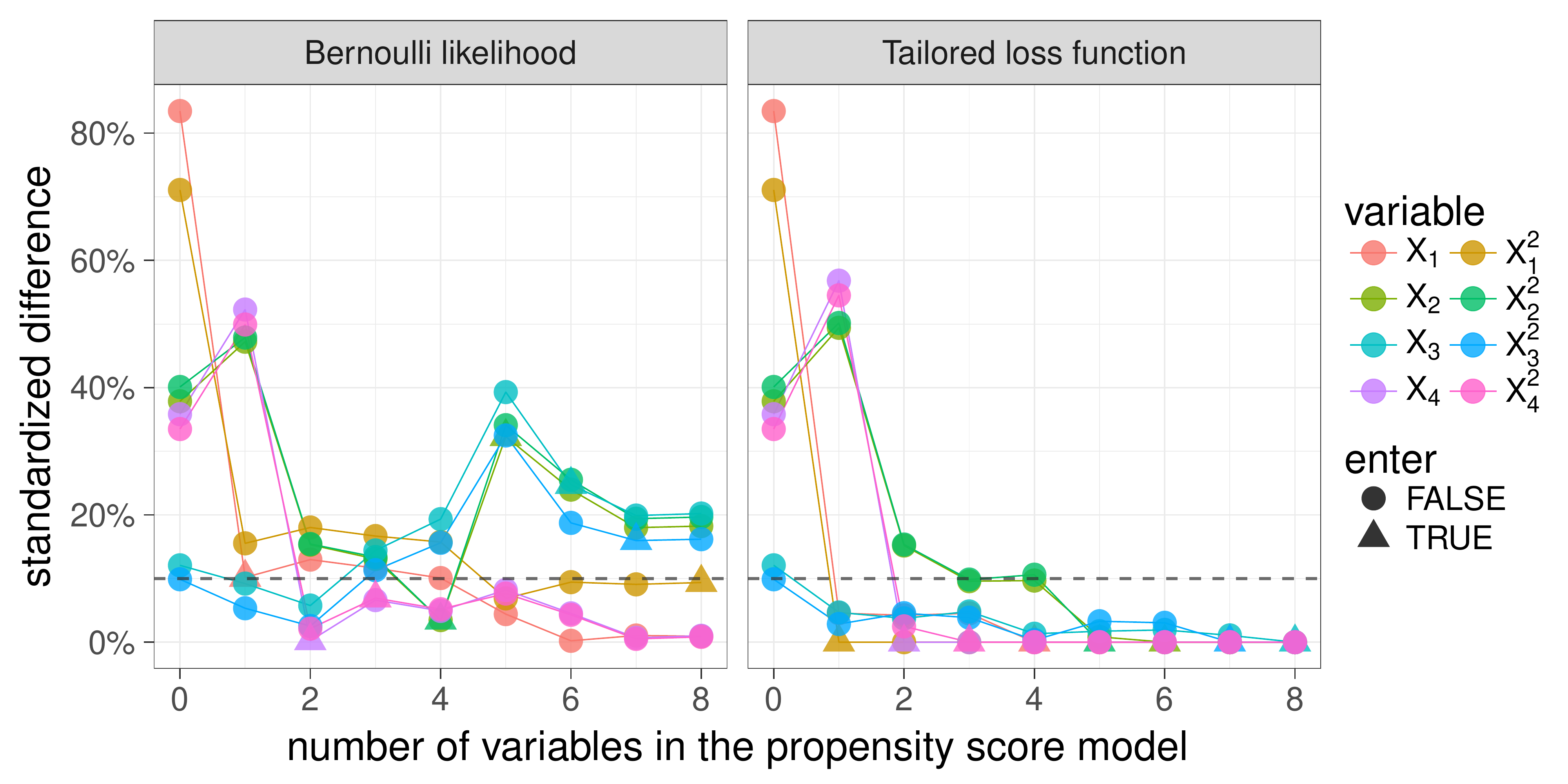}
  \caption{The tailored loss function proposed in
    this paper is much better than Bernoulli likelihood at reducing
    covariate imbalance. Propensity score is modeled by logistic
    regression and fitted by the tailored loss function or Bernoulli likelihood.
    Standardized difference is computed using inverse probability
    weighting (IPW) and pooled variance for the two treatment groups
    \citep{rosenbaum1985constructing}. A standardized
    difference above 10\% is often viewed unacceptable by many
    practitioners.}
  \label{fig:fstep-imba}
\end{figure}

\Cref{fig:fstep-imba} shows the paths of standardized difference for
one realization of the simulation. A widely used
criterion is that a standardized difference above 10\% is
unacceptable \citep{austin2015moving,normand2001validating}, which is
the dashed line in \Cref{fig:fstep-imba}. The left panel of
\Cref{fig:fstep-imba} uses the Bernoulli likelihood to fit and select
logistic regression models. The standardized difference paths are not
monotonically decreasing and never achieve the satisfactory level (10\%)
for all the regressors. In contrast, the right panel of
\Cref{fig:fstep-imba} uses the tailored loss function and all $8$
predictors are well balanced after $3$ steps. In fact, as a feature of
using the tailored loss function, all active regressors (variables in
the selected model) are exactly balanced.

The toy example here is merely for presentation, but it clearly
demonstrates that the proposed tailored loss function approach excels
in balancing covariates. We will discuss some practical strategies that are
more sophisticated than the forward stepwise regression in
\Cref{sec:extensions}.

\subsection{Related work and our contribution}
\label{sec:discussion}

The tailored loss function framework introduced here unifies a number
of existing methods by exploring the (Lagrangian) duality of
propensity scores and sample weights. Roughly speaking, the
``moment condition''
approaches advocated by \citet{graham2012inverse} and \citet{Imai2014} correspond
to the primal problem
of minimizing the tailored loss over propensity score models, while the
``empirical balancing'' proposals
\citep[e.g.][]{Hainmueller2011,zubizarreta2015stable} correspond to
the dual problem that solves some convex optimization problem over the
sample weights subject to covariate balance constraints.
The framework presented here is largely motivated by the
aforementioned works.
Part of the contribution of this paper is to bring together many
pieces scattered in this literature---moment condition of estimating
the propensity score, covariate balance, bias-variance trade-off,
different estimands, link function of a generalized linear model (the
latter two are often
overlooked)---and elucidate their roles
in the design and analysis of an observational study.

A reader familiar with the development of this literature may
recognize that many elements in the framework proposed here
have already appeared in some previous works. Perhaps the closest
approach is the covariate balancing propensity score method of
\citet{Imai2014}, as their covariate balancing moment conditions
are essentially the first-order conditions of minimizing the
tailored loss function. In fact, this is the reason that ``covariate
balancing propensity score'' is kept in the title of this
paper. However, by taking a loss-function based approach, we can
\begin{itemize}
\item Visualize the tailored loss functions which penalize more
  heavily on larger inverse probability weights and hence generates
  more stable estimates. (See the supplementary file for more detail.)
\item Understand when the moment equations have a unique solution by
  investigating convexity of the loss function. This may limit the
  estimands and link functions one can use in practice, though it turns out
  that for the logistic link function most of the commonly used estimands
  correspond to convex problems.
\item More importantly, allow the usage of predictive algorithms
  developed in statistical learning to optimize covariate balance in
  high-dimensional problems and rich function classes. Moment
  constraints methods usually exactly balance several selected
  covariate functions but leave the others unattended. By
  regularizing the tailored loss function, a signature technique in
  predictive modeling, the methods proposed in \Cref{sec:extensions} can
  inexactly balance high-dimensional or even infinite-dimensional
  covariate functions. This usually results in more accurate estimates
  of the weighted average treatment effects (WATE) and more robust
  statistical inference. 
\end{itemize}

Compared to the empirical balancing methods, the tailored loss function
framework shows that they are essentially equivalent to certain models of
propensity score. Asymptotic theory that are already established to
propensity-score based estimators can now apply to empirical balancing
methods. Our framework also allows the use of balancing weights in
estimating more general estimands. For example, we can produce balancing
weights to estimate the optimally weighted average treatment effect
proposed by \citet{crump2006moving} that is more stable when there is
limited overlap \citep{li2016balancing}.

Last but not the least, we provide a novel approach to make honest, design-based and
finite-sample inference for the weighted average treatment
effects (WATE). Instead of the improbable but commonly required assumption
that the propensity score is correctly specified, the only major
assumption we make is that the (unkonwn) true outcome regression function is in
a given class. The function class can be high-dimensional and
very rich. We give a Bayesian interpretation that underlies any
design of an observational study and provide extensive numerical
results to demonstrate the trade-off in making different assumptions
about the outcome regression function.

The next two Sections are devoted to introducing the tailored loss
functions. \Cref{sec:extensions} propose practical strategies
motivated by statistical learning. \Cref{sec:theoretical-aspects} then
considers some theoretical aspects about the tailored loss functions.
\Cref{sec:numerical-examples} uses numerical
examples in two new settings to demonstrate the flexibility of the
proposed framework and examine its empirical
performance. \Cref{sec:conclusions} concludes the paper with some
practical recommendations. Technical proofs are provided in the
supplementary file.


\section{Preliminaries on Statistical Decision Theory}
\label{sec:backgr-prop-scor}

To start with, propensity score estimation can be viewed as a decision
problem and this Section introduce some terminologies in statistical
decision theory. In a typical problem of making probabilistic
forecast, the decision maker needs to pick an element as the
prediction from $\mathcal{P}$, a convex class of probability measures
on some general sample space ${\Omega}$. For example, a weather
forecaster needs to report the chance of rain tomorrow, so the sample
space is $\Omega = \{\mathrm{rain}, \mathrm{no~rain}\}$ and the prediction is a
Bernoulli distribution.
Propensity score is a (conditional) probability measure, but recall
that the goal is to achieve satisfactory covariate balance rather than
the best prediction of treatment assignment. At a high level, this is
precisely the reason why we want to tailor the loss function when
estimating the propensity score.

\subsection{Proper scoring rules}
\label{sec:proper-scoring-rules}

At the core of statistical decision theory is the {\it scoring
  rule}, which can be any extended
real-valued function $S:\mathcal{P} \times \Omega \to
[-\infty,\infty]$ such that $S(P,\cdot)$ is
$\mathcal{P}$-integrable for all $P \in \mathcal{P}$ \citep{gneiting2007strictly}. If
the decision is $P$ and $\omega$ materializes, the decision maker's reward
or utility is $S(P,\omega)$. An equivalent but more pessimistic
terminology is \emph{loss function}, which is just the negative
scoring rule. These two terms will be used interchangeably in this
paper.

If the outcome is
probabilistic in nature and the actual probability distribution is
$Q$, the expected score of forecasting $P$ is
\begin{equation*} \label{eq:scoring-rule}
S(P,Q) = \int S(P, \omega) Q(d \omega).
\end{equation*}
To encourage honest decisions, the scoring rule $S$ is generally
required to be {\it proper},
\begin{equation}
  \label{eq:proper-rule}
  S(Q,Q) \ge S(P,Q),\quad \forall P,Q \in \mathcal{P}.
\end{equation}
The rule is called {\it strictly proper} if \eqref{eq:proper-rule}
holds with equality if and only if $P = Q$.

In observational studies, the sample space is commonly dichotomous
$\Omega = \{0,1\}$ (two treatment groups: $0$ for control and $1$ for
treated), though there is no essential difficulty to extend the
approach in this paper to $|\Omega| > 2$ (multiple treatments) or
$\Omega \subset \mathbb{R}$
(continuous treatment). In the binary case, a probability distribution
$P$ can be characterized by a single parameter $0\le p \le 1$, the
probability of treatment. A classical result of
\citet{savage1971elicitation} asserts that every real-valued (except
for possibly $S(0,1) = \infty$ or $S(1,0) = -\infty$) proper scoring
rule $S$ can be written as
\begin{equation*}
  \label{eq:proper-rule-binary}
  \begin{split}
    S(p,1) &= G(p) + (1-p) G'(p) = \int (1-p) G''(p) dp + \mathrm{const}, \\
    S(p,0) &= G(p) - p G'(p) = - \int p G''(p) dp + \mathrm{const}, \\
  \end{split}
\end{equation*}
where $G:[0,1] \to \mathbb{R}$ is a convex function and $G'(p)$ is a
subgradient of $G$ at the point $p \in [0,1]$. When $G$ is
second-order differentiable, an equivalent but more convenient
representation is
\begin{equation} \label{eq:proper-rule-deriv}
\frac{\partial}{\partial p} S(p,t) = (t - p) G''(p),~t=0,1.
\end{equation}
Since the function $G$ uniquely defines a scoring rule $S$, we shall
call $G$ a scoring rule as well.

A useful class of proper scoring rules is the following Beta family
\begin{equation}
  \label{eq:beta-family}
  G_{\alpha,\beta}''(p) = p^{\alpha - 1} (1-p)^{\beta - 1},~ -\infty < \alpha,\beta < \infty.
\end{equation}
These scoring rules were first introduced by \citet{buja2005loss} to
approximate the weighted misclassification loss by taking the limit
$\alpha,\beta \to \infty$ and $\alpha / \beta \to c$.  For example, if
$c = 1$, the score $G_{\alpha,\beta}$ converges to the zero-one
misclassification loss. Many important scoring rules belong to this family.
For example, the Bernoulli log-likelihood function $S(p,t) = t \log p + (1 - t) \log
(1-p)$ corresponds to $\alpha = \beta = 0$, and the Brier
score or the squared error loss
$S(p,t) = -(t - p)^2$ corresponds to $\alpha = \beta = 1$.
For our purpose of estimating propensity score, it will be shown later
that the subfamily $-1 \le \alpha,\beta \le 0$ is particularly useful.

\subsection{Propensity score modeling by maximizing score}
\label{sec:prop-score-estim}
Given i.i.d.\ observations $(X_i,T_i) \in \mathbb{R}^d \times
\{0,1\},~i=1,2,\dotsc,n$ where
$T_i$ is the binary treatment assignment and $X_i$ is a vector of $d$
pre-treatment covariates, we want to fit a model for the propensity
score $p(X) = \mathrm{P}(T = 1|X)$ in a prespecified family $\mathcal{P} =
\{p_{\theta}(X): \theta \in \Theta \}$. Later on we will consider very
rich model family, but for now let's focus on the generalized linear
models with finite-dimensional regressors $\phi(X) =
(\phi_1(X),\dotsc,\phi_m(X))^T$ \citep{mccullagh1989generalized}
\begin{equation} \label{eq:glm}
  p_{\theta}(X) = l^{-1}(f_{\theta}(X)) = l^{-1}(\theta^T \phi(X)),
\end{equation}
where $l$ is the \emph{link function}. In our framework, the tailored
loss function is determined by the link function $l$ (and the
estimand). The most common choice is the logistic link
\begin{equation} \label{eq:logistic-link}
  l(p) = \log \frac{p}{1-p},~l^{-1}(f) = \frac{e^f}{1 + e^f},
\end{equation}
which will be used in all the numerical examples of this paper.

Given a strictly proper
scoring rule $S$, the {\it maximum score (minimum loss)
  estimator} of $\theta$ is obtained by maximizing the average score
\begin{equation}
  \label{eq:opt-score-est}
\hat{\theta}_n = \arg \max_{\theta} \mathcal{S}_n(\theta) = \frac{1}{n} \sum_{i=1}^n
S(p_{\theta}(X_i),T_i).
\end{equation}
Notice that
an affine transformation $S(p,t) \mapsto a S(p,t) + b(t)$ for any $a
> 0$ and $-\infty < b(t) < \infty$ gives the same estimator
$\hat{\theta}_n$. Due to this reason, we will not differentiate between these
equivalent scoring rules and use a single function $S(p,t)$ to
represent all the equivalent ones.


When $S$ is differentiable and assuming exchangeability of taking
expectation and derivative, the maximizer of
$\mathrm{E}[\mathcal{S}_n(\theta)]$, which is indeed $\theta$ if the
propensity score is correctly specified $p(x)
= p_{\theta}(x)$ (a property called Fisher-consistency), is
characterized by the following estimating equations
\begin{equation} \label{eq:stationary-point}
  \nabla_{\theta} \, \mathrm{E}[\mathcal{S}_n(\theta)] =
  \mathrm{E}[\nabla_{\theta} \, \mathcal{S}_n(\theta)]
  =
  \mathrm{E}_{X,T}[\nabla_{\theta} \,
  S(l^{-1}(\theta^T\phi(X)),T)] = 0.
\end{equation}


\section{Tailoring the loss function}
\label{sec:tail-scor-rules}

\subsection{Covariate balancing scoring rules}
\label{sec:covar-balanc-scor}

The tailored loss function framework is motivated by reinterpreting
the first-order conditions \eqref{eq:stationary-point} as covariate
balancing constraints. Using the representation \eqref{eq:proper-rule-deriv} and the inverse
function theorem, we can rewrite \eqref{eq:stationary-point} as
\begin{equation}
  \label{eq:w-cov-bal-pop}
  \mathrm{E}[(T - (1 - T)) w(X,T)
  \cdot \phi(X)] = 0,
\end{equation}
where the weighting function
\begin{equation} \label{eq:gipw}
  w(x,t) = \frac{G''(p(x))}{l'(p(x))} [t (1 - p(x))
  + (1-t) p(x)]
\end{equation}
is determined by the scoring rule through $G''$ and the link function $l$.
The maximum score estimator $\hat{\theta}_n$ can be obtained by solving
\eqref{eq:w-cov-bal-pop} with the expectation over the
empirical distribution of $(X,T)$ instead of the population. When the
optimization problem \eqref{eq:opt-score-est} is strongly convex, the solution
to \eqref{eq:w-cov-bal-pop} is also unique.

The next key observation is that every weighting function $w(x,t)$ defines a
weighted average treatment effect (WATE). To see this, we
need to introduce the Neyman-Rubin causal
model. Let $Y(t),~t=0,1$ be the potential outcomes and $Y = TY(1) +
(1-T)Y(0)$ be the observed outcome. This paper assumes
strong ignorability of treatment assignment
\citep{rosenbaum1983}, so the observational study is free of hidden bias:
\begin{assumption} \label{assump:strong-ignor}
  $T \independent (Y(0),Y(1)) | X$.
\end{assumption}

Let the observed outcomes be $Y_i$, $i=1,\dotsc,n$. Naturally, the
weighted difference of $Y_i$,
\begin{equation} \label{eq:tau-hat}
  \hat{\tau} = \sum_{i:\,T_i = 1} w(X_i,T_i) Y_i - \sum_{i:\,T_i = 0} w(X_i,T_i) Y_i,
\end{equation}
estimates the following population parameter
\[
\tau_v = \mathrm{E}_{X,T,Y}\{(T - (1 - T)) w(X,T) Y\} = \mathrm{E}_{X,Y} \left[v(X) (Y(1) - Y(0))\right],
\]
which is an (unnormalized) weighted average treatment effect. Here
\begin{equation} \label{eq:w-g}
v(X) = \mathrm{E}[T \cdot w(X,1) | X] = \mathrm{E}[(1-T) \cdot w(X,0)|X] = \frac{G''(p(X)) \, p(X) \, (1 - p(X))}{l'(p(X))}.
\end{equation}
In practice, it is usually more meaningful to estimate the
normalized version $\tau_w^{*} = \tau_w / \mathrm{E}\left[w(X)\right]$
by normalizing the weights $w_i = w(X_i,T_i),~i=1,\dotsc,n$ separately among
the treated and the control: $\hat{w}_i^{*} = \hat{w}_i /
\sum_{j:\,T_j=T_i} \hat{w}_j$, $i=1,\dotsc,n$.

\Cref{tab:scoring-rule} shows that four mostly commonly used estimands, the average
treatment effect (ATE), the average treatment effect on the control
(ATC), the average treatment effect on the treated (ATT), and the
optimally weighted average treatment effect (OWATE) under homoscedasticity
\citep{crump2006moving}, are weighted average
treatment effects with
\begin{equation} \label{eq:w-alpha-beta}
v_{\alpha,\beta}(X) = p({X})^{\alpha+1} (1 - p({X}))^{\beta+1}
\end{equation}
with different combinations of $(\alpha,\beta)$.

\def\arraystretch{1.8}
\renewcommand{\tabcolsep}{2pt}
\begin{table}[t]
  \centering
  \caption{Correspondence of estimands, sample weighting functions,
    and the covariate balancing scoring rules (corresponding to the
    logistic link) in the proposed Beta family. The estimand is a weighted
    average treatment effect $\tau_{\alpha,\beta} = \mathrm{E} \left[v_{\alpha,\beta}(X)
      (Y(1) - Y(0))\right]$ and $\tau^{*} = \tau_{\alpha,\beta} /
    \mathrm{E}[v_{\alpha,\beta}(X)]$.}
  \label{tab:scoring-rule}
  \small
  \begin{tabular}{|c|c|c|c|c|c|c|}
    \hline
    ~~$\alpha$~~ & ~~$\beta$~~ & estimand  &
    $w(x, 1)$ & $w(x, 0)$ & $S(p,1)$ & $S(p, 0)$\\
    \hline
    -1 & -1 & $\tau = \tau^{*} = \mathrm{E}[Y(1) - Y(0)]$ & $\frac{1}{p(x)}$ & $\frac{1}{1-p(x)}$ & $\log \frac{p}{1-p} - \frac{1}{p}$
    & $ \log \frac{1-p}{p} - \frac{1}{1-p}$\\
    \hline
    -1 & 0 & $\tau^{*} = \mathrm{E}[Y(1) - Y(0)|T=0]$ & $\frac{1-p(x)}{p(x)}$ & $1$ & $-\frac{1}{p}$
    & $ \log \frac{1-p}{p}$\\
    \hline
    0 & -1 & $\tau^{*} = \mathrm{E}[Y(1) - Y(0)|T=1]$  & $1$ & $\frac{p(x)}{1-p(x)}$ & $ \log
    \frac{p}{1-p}$ & $-\frac{1}{1-p}$\\
    \hline
    0 & 0 & \scriptsize $\tau = \mathrm{E}[p(X) (1-p(X)) \cdot (Y(1) - Y(0))]$ & $1 - p(x)$ & $p(x)$ &
    $\log p$ & $\log(1-p)$ \\
    \hline
  \end{tabular}
\end{table}

Therefore, in order to estimate $\tau_{\alpha,\beta} =
\mathrm{E}[v_{\alpha,\beta}(X)(Y(1)-Y(0))]$, we just need to equate
\eqref{eq:w-g} with \eqref{eq:w-alpha-beta} and solve for $G$. The
solution in general depends on the link function $l$. If the logistic
link is used, it is easy to show that the solution belongs to
the Beta family of scoring rules defined in
\eqref{eq:beta-family}. The loss functions corresponding to the four
estimands are also listed in \eqref{eq:w-alpha-beta}.
\begin{proposition} \label{prop:correspondence}
  Under \Cref{assump:strong-ignor}, if $l$ is the logistic link
  function, then $\tau_v = \tau_{\alpha,\beta}$ if $G =
  G_{\alpha,\beta}$.
\end{proposition}

To use the framework developed here in practice, the user should ``invert'' the development in this Section. First, the user should determine the estimand by its interpretation
and whether there is insufficient covariate overlap (so OWATE may be
desirable). Second, the user should decide on a link function (we
recommend logistic link). Lastly, the user can equate \eqref{eq:w-g} with
\eqref{eq:w-alpha-beta} or look up
\Cref{tab:scoring-rule} to find the corresponding scoring rule.

The main advantage of using the ``correct'' scoring rule is that the
weights will automatically balance the predictors $\phi(X)$. This is a
direct consequence of the estimating equations
\eqref{eq:w-cov-bal-pop} and is summarized in the next theorem. This is
precisely the reason we call $G_{\alpha,\beta}$ or the corresponding
$S_{\alpha,\beta}$ the {\it covariate balancing scoring rule} (CBSR) with
respect to the estimand $\tau_{\alpha,\beta}$ and the logistic link
function in this paper.

\begin{theorem} \label{thm:covariate-balancing}
  If $l$ is the logistic link function and
  $\hat{\theta}$ is obtained as in
  \eqref{eq:opt-score-est} by maximizing the CBSR corresponding to
  $l$ and the estimand. Then the weights
  $\hat{w}_i,~i=1,\dotsc,n$ computed by \eqref{eq:gipw} exactly
  balance the sample regressors
  \begin{equation}
    \label{eq:w-cov-bal}
    \sum_{i:\,T_i = 1} \hat{w}_i \phi(X_i) = \sum_{i:\,T_i = 0} \hat{w}_i \phi(X_i).
  \end{equation}
  Furthermore, if the predictors include an intercept term (i.e. $1$ is
  in the linear span of $\phi(X)$), then $ \hat{w}^{*}$ also satisfies
  \eqref{eq:w-cov-bal}.
\end{theorem}

Note that the Bernoulli likelihood
($\alpha = \beta = 0$) indeed corresponds to the estimand OWATE
instead of the more commonly used ATE or ATT. This corresponds to the
``overlap weights'' recently proposed by \citet{li2016balancing},
where each observation's weight is proportional to the probability of
being assigned to the opposite group. Theorem 3 of \citet{li2016balancing}
states that the ``overlap weights'' exactly balances the regressors
when Bernoulli likelihood is used, which is a special case of our
\Cref{thm:covariate-balancing}.

\subsection{Convexity}
\label{sec:convexity}

To obtain covariate balancing propensity scores, one can solve the
estimating equations \eqref{eq:w-cov-bal-pop} directly without using
the tailored loss function. This is essentially the approach
taken by \citet{Imai2014}, although it is unclear at this point that
\eqref{eq:w-cov-bal} has a unique solution. The first advantage of
introducing the tailored loss functions is that some CBSR is strongly
concave, so the solution to its first-order condition is always unique.

\begin{proposition} \label{prop:beta-concave}
  Suppose the estimand is in the Beta-family \Cref{eq:w-alpha-beta}
  and let $S$ be the CBSR corresponding to a link function $l$ such
  that $v = v_{\alpha,\beta}$. Then the score functions
  $S(l^{-1}(f),0)$ and $S(l^{-1}(f),1)$ are both concave functions of $f \in
  \mathbb{R}$ if and only if $-1 \le \alpha, \beta \le 1$. Moreover,
  if $(\alpha,\beta) \ne (-1,0)$,
  $S(l^{-1}(f),0)$ is strongly concave; if $(\alpha,\beta) \ne
  (0,-1)$, $S(l^{-1}(f),1)$ is strongly concave.
\end{proposition}

Notice that the range of $(\alpha,\beta)$ in \Cref{prop:beta-concave}
includes the four estimands listed in \Cref{tab:scoring-rule}.
As a consequence, their corresponding score maximization problems can
be solved very efficiently (for example by Newton's method).
Motivated by this observation, in the next
Section we propose to fit propensity score models with more
sophisiticated strategies stemming from statistical learning.


\section{Adaptive Strategies}
\label{sec:extensions}

The generalized linear model considered in \Cref{sec:tail-scor-rules}
amounts to a fixed low-dimensional model space
\[
  \mathcal{P}_{\mathrm{GLM}} = \Big\{p(x) = l^{-1}(f(x)): \,
  f(x) \in \mathrm{span}(\phi_1(x),\phi_2(x),\dotsc,\phi_m(x))^T\Big\}.
\]

As mentioned previously in \Cref{sec:introduction}, in principle, we
should not restrict to a single propensity score model as it can be
misspecified. Propensity score is merely a nuisance parameter in estimating
WATE. We shall see repeatedly in later Sections
that, in finite sample, it is more important to
use flexible propensity score models that balance covariates well than
to estimate the propensity score accurately. In this Section, we
incorporate machine learning methods in our framework to expand the
model space.

\subsection{Forward stepwise regression}
\label{sec:forward-stepwise}

To increase model complexity, perhaps the most straightforward
approach is forward stepwise regression as illustrated earlier in
\Cref{sec:introduction}. Instead of a fixed model space, forward
stepwise gradually increases model complexity. Using the tailored loss
functions in \Cref{sec:tail-scor-rules}, active covariates are always
exactly balanced and inactive covariates are usually well balanced too.

Motivated by this strategy, \citet*{hirano2003} studied the efficiency
of the IPW estimator when the dimension of the regressors $\phi(x)$ is
allowed to increase as the sample size $n$ grows. Their renowned
results claim that this \emph{sieve} IPW estimator is
semiparametrically efficient for estimating the WATE. Here we show that the semiparametric efficiency
still holds if the Bernoulli likelihood, the loss function that
\citet{hirano2003} used to estimate the propensity score, is
replaced by the Beta family of scoring rules $G_{\alpha,\beta}$, $-1
\le \alpha,\beta \le 0$ in \eqref{eq:beta-family} or essentially any
strongly concave scoring rule. This result is not too
surprising as the propensity score is just a nuisance parameter
whose estimation accuracy is of less importance in semiparametric
inference. Conceptually, however, this result suggests that the
investigator has the freedom to choose the loss function in
estimating the propensity score and do not need to worry about loss of
asymptotic efficiency. The advantages of using a tailored loss function
are better accuracy in finite sample and more robustness against model
misspecification, as detailed later in the
\Cref{sec:theoretical-aspects}.

Let's briefly review the sieve logistic regression in
\citet{hirano2003}. For $m = 1,2,\dotsc$, let ${\phi}_m({x}) =
(\varphi_{1m}({x}),\varphi_{2m}({x}),\dotsc,\varphi_{mm}({x}))^T$ be a
triangular array of orthogonal polynomials, which are obtained by orthogonalizing the power series:
$\psi_{km}({x}) = \prod_{j=1}^d x_j^{\gamma_{kj}}$,
where ${\gamma}_k = (\gamma_{k1},\dotsc,\gamma_{kd})^T$ is an
$d$-dimensional multi-index of nonnegative integers and satisfies
$\sum_{j=1}^d \gamma_{kj} \le \sum_{j=1}^d \gamma_{k+1,j}$.
Let $l$ be the logistic link function \eqref{eq:logistic-link}.
\citet{hirano2003} estimated the propensity score by maximizing the log-likelihood
\[
\hat{{\theta}}^{\mathrm{MLE}} = \arg \max_{{\theta}}
\frac{1}{n}\sum_{i=1}^n T_i \log \left( l^{-1}({\phi}_m({X}_i)^T {\theta})\right) + (1 -
T_i) \log \left(1 - l^{-1}({\phi}_m({X}_i)^T {\theta})\right).
\]
This is a special case of the score maximization problem
\eqref{eq:opt-score-est} by setting $S = S_{0,0}$.

\Cref{thm:beta-efficient} below is an extension to the main theorem of
\citet{hirano2003}. Besides strong ignorability, the other
technical assumptions in \citet{hirano2003} are listed in the
supplement. Compared to the original theorem which always uses
the maximum likelihood regardless of the estimand, the
scoring rule is now tailored according to the estimand as described in
\Cref{sec:tail-scor-rules}.

\begin{theorem} \label{thm:beta-efficient}
Suppose we use the Beta-family of
  covariate balancing scoring rules defined by
  \cref{eq:proper-rule-deriv,eq:beta-family} with $-1 \le \alpha,\beta
  \le 0$ and the logistic link \eqref{eq:logistic-link}.
  Under \Cref{assump:strong-ignor} and the technical assumptions in
  \citet{hirano2003}, if we choose suitable $m$ growing with $n$, the propensity score
  weighting estimator $\hat{\tau}_{\alpha,\beta}$ and its normalized version $\hat{\tau}_{\alpha,\beta}^{*}$ are
  consistent for $\tau_{\alpha,\beta}$ and $\tau_{\alpha,\beta}^{*}$. Moreover, they reach the semiparametric
  efficiency bound for estimating $\tau_{\alpha,\beta}$ and $\tau_{\alpha,\beta}^{*}$.
\end{theorem}

\subsection{Regularized Regression}
\label{sec:regul-regr}

In predictive modeling, stepwise regression is usually sub-optimal
especially if we have high dimensional covariates \citep[see e.g.][Section
3]{hastie2009elements}. A more principled approach is to penalize the
loss function
\begin{equation}
  \label{eq:regularized-glm}
  \hat{\theta}_{\lambda} = \arg \max_{p(\cdot) \in \mathcal{P}} ~ \frac{1}{n} \sum_{i=1}^n
  S(p({X}_i), T_i) - \lambda J(p(\cdot)),~p(x) = l^{-1}(f(x)),
\end{equation}
where $J(\cdot)$ is a regularization function that penalizes
overly-complicated propensity score model $p(x)$ and the tuning
parameter $\lambda$
controls the degree of regularization. This estimator reduces to the
optimum score estimator \eqref{eq:opt-score-est} when $\lambda = 0$.

The penalty term $J(\theta)$ should be chosen according to the
model space $\mathcal{P}$ and the investigator's prior belief about
the outcome regression function (see \Cref{sec:bayes-interpr}).
In this paper, we consider three alternatives of model space and penalty:

\begin{enumerate}
\item Regularized GLM: the model space is the same generalized linear
  model $\mathcal{P}_{\mathrm{GLM}}$ with potentially
  high dimensional covariates, but the average score is penality by
  the $l_a$-norm of $\theta$, $J(p_{\theta}) =
  \frac{1}{a} \sum_{k=1}^m |\theta_k|^a$ for some $a \ge 1$.
  Some typical choices are the $l_1$ norm
  $J(p_{\theta}) = \|{\theta}\|_1$ (lasso) and the squared $l_{2}$ norm $J(p_{\theta}) =
  \|{\theta}\|_2^2$ (ridge regression).
\item Reproducing kernel Hilbert space (RKHS): the model space is the
  RKHS generated by a given kernel $K$, $\mathcal{P}_{\mathrm{RKHS}} =
  l^{-1}(\mathcal{H}_K)$, and the penalty is the corresponding norm of
  $f$, $J(p(\cdot)) = \|f\|_{\mathcal{H}_K}$.
\item Boosted trees: the model space is the additive trees:
  $\mathcal{P}_{k\textrm{-}\mathrm{boost}} = \{ p(x) = l^{-1}(f(x)): f
  = f_1 + f_2 +
  \cdots + f_m: f_k \in
  \mathcal{F}_{\mathrm{d\textrm{-tree}}},~k=1,2,\dotsc\}$, where
  $\mathcal{F}_{\mathrm{d\textrm{-tree}}}$ is the space of
  step functions in the classification and regression tree (CART) with depth at
  most $d$ \citep{breiman1984classification}. This space is quite
  large and approximate fitting algorithms (boosting) must be
  used. There is no exact penalty function, but as noticed by
  \citet{Friedman1998} and illustrated later, boosting is closely
  related to the lasso penalty in regularized regression.
\end{enumerate}

Since all the penalty terms considered here are conex, the regularized
optimization problems can be solved very efficiently.

\subsection{RKHS regression}
\label{sec:rkhs}

Next, we elaborate on the RKHS and boosting approaches since they might
be foreign to researchers in causal inference. RKHS regression is a
popular nonparametric method in machine learning that
essentially extends the regularized GLM with ridge penalty to an
infinite dimensional space
\citep{wahba1990spline,hofmann2008kernel}. Let $\mathcal{H}_K$ be the
RKHS generated by the kernel function $K:\mathcal{X} \times
\mathcal{X} \to \mathbb{R}$, which describes similarity between
two vectors of pre-treatment covariates. The RKHS model is most easily
understood through the ``feature map'' interpretation.
Suppose that $K$ has an eigen-expansion $K({x},{x}') = \sum_{k=1}^{\infty} c_k \phi_k({x}) \phi_k({x}')$
with $c_k \ge 0,~\sum_{k=1}^{\infty} c_k^2 < \infty$. Elements of
$\mathcal{H}_K$ have a series expansion
\[
  f({x}) = \sum_{k=1}^{\infty} \theta_k
  \phi_k({x}),~\|f\|_{\mathcal{H}_K}^2 = \sum_{k=1}^{\infty} \theta_k^2/c_k.
\]
The eigen-functions $\{\phi_1(x),\phi_2(x),\dotsc,\}$ can be viewed as
new regressors generated by the low-dimensional covariates $X$.
The standard generalized linear model \eqref{eq:glm} corresponds to a
finite-dimensional linear
reproducing kernel $K({x},{x}') = \sum_{k=1}^m \phi_k({x})
\phi_k({x}')$, but in general the eigen-functions (i.e.\ predictors)
$\{\phi_k\}_{k=1}^{\infty}$ can be infinite-dimensional.

Although the parameter $\theta$ is potentially infinite-dimensional, the
numerical problem \eqref{eq:regularized-glm} is computationally
feasible via the ``kernel trick'' if the penalty is a function of the
RKHS norm of $f(\cdot)$. The representer theorem
\citep[c.f.][]{wahba1990spline} states that the solution is indeed
finite-dimensional and has
the form $\hat{f}({x}) = \sum_{i=1}^n \hat{\gamma}_i K({x},
{X}_i)$. Consequently, the optimization problem \eqref{eq:regularized-glm} can
be solved with the $n$-dimension parameter vector $\gamma$.

As a remark, the idea of using a kernel to describe similarity between
covariate vectors is not entirely new to observational studies.
However, most of the previous literature
\citep[e.g.][]{heckman1997matching} uses kernel as
a smoothing technique for propensity score estimation (similar to
kernel density estimation) rather than
generating a RKHS, although the kernel functions can be the same in
principle.

\subsection{Boosting}
\label{sec:boosting}

Boosting (particularly gradient boosting) can be viewed as a greedy
algorithm of function approximation \citep{friedman2001greedy}. Let
$\hat{f}$ be the current guess of $f$, then the next guess is given by
the steepest gradient descent $\hat{f}_{\mathrm{new}} = \hat{f} +
\hat{\eta} \hat{h}$, where
\begin{align}
  \label{eq:g-grad}
  \hat{h} &= \arg\max_{h \in \mathcal{F}_{k\textrm{-tree}}} \frac{\partial}{\partial \eta}
  \frac{1}{n} \sum_{i=1}^n S_{\alpha,\beta}\left(l^{-1}(\hat{f}(X_i) + \eta
  h(X_i)), T_i\right),~\mathrm{and} \\
  \hat{\eta} &= \arg\max_{\eta \ge 0}   \frac{1}{n} \sum_{i=1}^n
  S_{\alpha,\beta}\left(l^{-1}(\hat{f}(X_i) + \eta \hat{h}(X_i)), T_i\right).
\end{align}

When using gradient boosting in predictive modeling, a practical
advice is to not go fully along the gradient direction as it easily
overfits the model. \citet{friedman2001greedy} introduced an
tuning parameter $\nu > 0$ (usually much less than $1$) and proposed
to shrink each gradient update: $\hat{f}_{\mathrm{new}} = \hat{f} +
\nu \hat{\eta} \hat{h}$. Heuristically, this can be compared with the
difference between the forward stepwise regression which commits to the selected
variables fully and the least angle regression or the lasso regression which moves an
infinitesimal step forward each time \citep{efron2004least}. We shall
see in the next Section that, in the context of propensity score
estimation, boosting and lasso regression also share a similar dual
interpretation.

\subsection{Adjustment by outcome regression}
\label{sec:outcome-regression}

So far we have only considered design-based estimators by building a
propensity score model to weight the observations. Such estimators do
not attempt to build models for the potential outcomes, $Y(0)$ and
$Y(1)$. Design-based inference is arguably more straightforward as it
attempts to mimic a randomized experiment by observation
data. Nevertheless, in some applications it is reasonable to improve
estimation accuracy by fitting outcome regression models.

Here we describe the augmented inverse probability weighting (AIPW)
estimators of ATT and ATE. Let
$g_0(X) = \mathrm{E}[Y(0)|X]$ and $g_1(X) = \mathrm{E}[Y(1)|X]$ be the true
regression functions of the potential outcomes and $\hat{g}_0$ and
$\hat{g}_1$ be the corresponding estimates. Let $w^{\mathrm{ATT}}$ and
$w^{\mathrm{ATE}}$ be the weights obtained by maximizing CBSR
($S_{0,\textrm{-}1}$ for ATT and $S_{\textrm{-}1,\textrm{-}1}$ for
ATE) with any of the above adaptive strategies. The AIPW
estimators \citep{Robins1994} are
\[
\hat{\tau}_{\mathrm{AIPW}}^{\mathrm{ATT}} = \sum_{T_i=1}
w^{\mathrm{ATT}}_i (Y_i - \hat{g}_0(X_i)) - \sum_{T_i=0}
w^{\mathrm{ATT}}_i (Y_i - \hat{g}_0(X_i)),~\mathrm{and}
\]
\[
\hat{\tau}_{\mathrm{AIPW}}^{\mathrm{ATE}} = \frac{1}{n} \sum_{i=1}^{n}
(\hat{g}_1(X_i) - \hat{g}_0(X_i)) + \sum_{T_i=1}
w^{\mathrm{ATE}}_i (Y_i - \hat{g}_1(X_i)) - \sum_{T_i=0}
w^{\mathrm{ATE}}_i (Y_i - \hat{g}_0(X_i)).
\]

We will compare IPW and AIPW estimators in the numerical examples in
\Cref{sec:numerical-examples}.

\section{The Dual Intepretation}
\label{sec:theoretical-aspects}

We have proposed a very general framework and several flexible
methods to estimate the propensity score. Several important questions
are left unsettled: if different loss functions are asymptotically
equivalent as indicated by \Cref{thm:beta-efficient}, why should we
use the tailored loss functions in this paper (or any method listed in
\Cref{sec:discussion})? How should we choose among the adaptive
strategies in \Cref{sec:extensions}? What is the bias-variance
trade-off in regularizing the propensity score model and how should we
choose the regularization parameter $\lambda$ in
\Cref{eq:regularized-glm}? After fitting a propensity score model, how
do we construct a confidence interval for the target parameter? This
Section addresses these questions through investigating the
Lagrangian dual of the propensity score estimation problem.

\subsection{Covariate imbalance and bias}
\label{sec:impl-covar-balance}

As in \Cref{sec:outcome-regression}, denote the true outcome
regression functions by $g_t({X}) =
\mathrm{E}[Y(t)|{X}],~t=0,1$. Except for ATT, in
this Section we will only consider bias under the constant treatment
effect model that
$g_1(x) = g_0(x) + \tau^{*}$ for all $x$. By definition, $\tau^{*}$ is
also the (normalized) weighted average treatment effects.

Suppose $g_0(x)$ has the expansion $g_0(x) = \sum_{k=1}^{\infty} \beta_k
\phi_k(x)$ for some functions $\{\phi_1(x),\phi_2(x),\cdots\}$. Let
$\epsilon_i = Y_i - g_{T_i}(X_i)$ so $\mathrm{E}[\epsilon_i|T_i,X_i] =
0$. Given any weighting function $w(x,t),~t=0,1$ on the sample (e.g.\
\cref{eq:w-g,eq:w-alpha-beta} with estimated propensity score) and
denote $w_i = w(X_i,T_i)$, the IPW-type estimator
$\hat{\tau}^{*}$ defined in \eqref{eq:tau-hat} has the decomposition
\begin{equation} \label{eq:bias-tau}
  \begin{split}
    \hat{\tau}^{*} &= \tau^{*} +
    \Big[\sum_{T_i=1}
    w_i g_0(X_i) - \sum_{T_i=0} w_i g_0(X_i) \Big] + \Big[\sum_{T_i=1}
    w_i \epsilon_i - \sum_{T_i=0} w_i \epsilon_i \Big]\\
    &= \sum_{k=1}^{\infty} \beta_k \cdot \Big[\sum_{T_i=1}
    w_i \phi_k(X_i) - \sum_{T_i=0} w_i \phi_k(X_i) \Big] + \Big[\sum_{T_i=1}
    w_i \epsilon_i - \sum_{T_i=0} w_i \epsilon_i \Big]\\
  \end{split}
\end{equation}
The second term always has mean $0$, so the bias of $\hat{\tau}^{*}$ is
given by the first term (a fixed quantity conditional on
$\{T_i,X_i\}_{i=1}^n$), which is just the imbalance with respect to the
covariate function $g_0(x)$. The second line decomposes the bias into
the imbalance with respect to the basis functions
$\{\phi_1(x),\phi_2(x),\cdots\}$.

\Cref{eq:bias-tau} highlights the importance of covariate balance
in reducing the bias of $\hat{\tau}^{*}$, especially if the propensity
score model is misspecified. If the propensity score is modeled by a
GLM with fixed regressors $\phi(x) = (\phi_1(x),\dotsc,\phi_m(x))$ and fitted
by maximizing CBSR as in \eqref{eq:opt-score-est}, an immediate
corollary is
\begin{theorem} \label{prop:unbiased-global-null}
  Under \Cref{assump:strong-ignor} and constant treatment effect that
  $g_1(x) \equiv  g_0(x) + \tau^{*}$ for all
  ${x}$, the estimator $\hat{\tau}^{*}$ obtained by maximizing CBSR
  with regressors $\phi(x)$ that include an intercept term is asymptotically unbiased if
  $g_0({x})$ is in the linear span of
  $\{\phi_1({x}),\dotsc,\phi_m({x})\}$, or more generally if
  $\inf_{\eta}\|g_0(x) - \eta^T \phi_m(x)\|_{\infty} \to 0$ as $n,m(n)
  \to \infty$.
\end{theorem}

The last condition says that $g_0(x)$ can be uniformly approximated by
functions in the linear span of $\phi_1(x),\dotsc,\phi_m(x)$ as $m \to
\infty$. This holds under very mild assumption of $g_0$. For example, if
the support of $X$ is compact and $g_0(x)$ is continuous, the
Weierstrass approximation theorem ensures that $g_0(x)$ can be uniformly
approximated by polynomials.

Finally we compare the results in
\Cref{thm:beta-efficient,prop:unbiased-global-null}. The main
difference is that \Cref{thm:beta-efficient} uses
{\it propensity score} models with increasing complexity, whereas
\Cref{prop:unbiased-global-null} assumes uniform approximation for the
{\it outcome regression} function. Since the unbiasedness in
\Cref{prop:unbiased-global-null} does not make any assumption on
the propensity score, the estimator $\hat{\tau}^{*}$ obtained by
maximizing CBSR is more robust to misspecified or overfitted
propensity score model.

\subsection{Lagrangian Duality}
\label{sec:langrangian-duality}

In \Cref{sec:discussion}, we mentioned that the recently proposed ``moment
condition'' approaches \citep[e.g.][]{Imai2014} and the ``empirical
balancing'' approaches \citep[e.g.][]{zubizarreta2015stable} can be
unified under the framework proposed in this paper. We now elucidate
this equivalence by exploring the Lagrangian dual of maximizing CBSR.
First, let's rewrite the score optimization problem
\eqref{eq:opt-score-est} by introducing new variables $f_i$ for each
observation $i$:
\begin{equation}
  \label{eq:proper-rule-max-ps-eq}
  \begin{split}
    \underset{f,{\theta}}{\mathrm{maximize}} \quad& \frac{1}{n} \sum_{i=1}^n
    S(l^{-1}(f_i),T_i) \\
    \mathrm{subject~to} \quad& f_i = {\theta}^T {\phi}({X}_i),~i=1,\dotsc,n.
  \end{split}
\end{equation}
Let the Lagrangian multiplier associated with the $i$-th constraint be
$(2T_i - 1)w_i/n$.
By setting the partial derivatives of the Lagrangian equal to $0$, we obtain
\begin{align}
  \frac{\partial Lag}{\partial \theta_k} &= \frac{1}{n} \sum_{i=1}^n
  (2T_i - 1){w}_i \phi_k({X}_i) = 0,~k=1,\dotsc,m. \label{eq:kkt-1} \\
  \frac{\partial Lag}{\partial f_i} &= \frac{1}{n} \left( \frac{\partial
      S(l^{-1}(f_i), T_i)}{\partial f_i} + (2T_i - 1){w}_i \right) = 0,~i=1,\dotsc,n, \label{eq:kkt-2}
\end{align}
Equation \eqref{eq:kkt-1} is the same as \eqref{eq:w-cov-bal}, meaning
the optimal dual variables $w$ balance the predictors
$\phi_1,\dotsc,\phi_m$.
Equation \eqref{eq:kkt-2} determines $w$ from $f$. By using the fact
\eqref{eq:proper-rule-deriv} and the scoring rule is CBSR, it turns
out that $w_i = w(X_i,T_i)$ is
exactly the weights defined in \eqref{eq:gipw}, that is,
the weights $w$ in our estimator $\hat{\tau}$ are dual variables
of the CBSR-maximization problem \eqref{eq:opt-score-est}.

Next we write down the Lagrangian dual problem of
\eqref{eq:proper-rule-max-ps-eq}.
In general, there is no explicit form because it is difficult to invert
\eqref{eq:gipw}. However, in the particularly interesting cases
$\alpha =0, \beta = -1$ (ATT) and $\alpha = -1,\beta = -1$ (ATE), the
dual problems are algebraically tractable. When $\alpha = 0, \beta =
-1$ and if an intercept term is included in the GLM, 
(an equivalent form of) the dual problem is given by
\begin{equation}
  \label{eq:proper-rule-max-ps-att-dual}
  \begin{split}
    \underset{{w} \ge 0}{\mathrm{minimize}}\quad
    &\sum_{i:T_i=0} w_i \log w_i\\
    \mathrm{subject~to} \quad & \sum_{i:T_i=0} w_i \phi_k({X}_i) =
    \sum_{j:T_j = 1} \phi_k({X}_j),~k=1,\dotsc,m.
  \end{split}
\end{equation}

When $\alpha = \beta = -1$, the inverse probability weights are always
greater than $1$. The Lagrangian dual problem in
this case is given by
\begin{equation}
  \label{eq:proper-rule-max-ps-dual}
  \begin{split}
    {\mathrm{minimize}}\quad
    &\sum_{i=1}^n (w_i - 1) \log (w_i - 1) - w_i\\
    \mathrm{subject~to} \quad & \sum_{i:T_i=0} w_i \phi_k({X}_i) =
    \sum_{j:T_j = 1} w_j \phi_k({X}_j),~k=1,\dotsc,m. \\
    & w_i \ge 1,~i=1,\dotsc,n.
  \end{split}
\end{equation}

The objective functions in \eqref{eq:proper-rule-max-ps-att-dual} and
\eqref{eq:proper-rule-max-ps-dual} encourage the weights $w$ to be
close to uniform. They belong to a general distance measure $\sum_{i=1}^n
D(w_i,v_i)$ in \citet{deville1992calibration}, where $D(w,v)$
is a continuously differentiable and strongly convex
function in $w$ and achieves its minimum at $w = v$. When
the estimand is ATT (or ATE), the target weight $v$ is equal to $1$ (or
$2$). Estimators of this kind are often
called ``calibration estimators'' in survey sampling because the
weighted sample averages are empirically calibrated to some known
population averages.

All the previously proposed ``empirical balancing'' methods operate by
solving convex optimization problems similar to
\eqref{eq:proper-rule-max-ps-att-dual} or \eqref{eq:proper-rule-max-ps-dual}.
The maximum entropy problem \eqref{eq:proper-rule-max-ps-att-dual}
appeared first in \citet{Hainmueller2011} to estimate ATT and is called ``Entropy
Balancing''. \citet{zhao2015} used the primal-dual connection
described above to show Entropy Balancing is doubly robust, a stronger
property than \Cref{thm:beta-efficient,prop:unbiased-global-null}.
Unfortunately, the double robustness property does not extend to other estimands.
\citet{chan2015} studied the calibration
estimators with the general distance $D$ and showed the estimator
$\hat{\tau}$ is globally semiparametric efficient. When the estimand
is ATE, \citet{chan2015} require the weighted sums of $\phi_k$ in
\eqref{eq:proper-rule-max-ps-dual} to be calibrated to $\sum_{i=1}^n
\phi_k(X_i)/n$, too. In view of \Cref{thm:beta-efficient}, this extra
calibration is not necessary for semiparametric efficiency. In an
extension to Entropy Balancing, \citet{hazlett2013balancing} proposed
to empirically balance kernel representers instead of fixed
regressors. This corresponds to unregularized ($\lambda=0$) RKHS
regression introduced in \Cref{sec:rkhs}. The unregularized
problem is in general unfeasible, so \citet{hazlett2013balancing} had
to tweak the objective to find a usable
solution. \citet{zubizarreta2015stable} proposed to solve a problem
similar to \eqref{eq:proper-rule-max-ps-dual} (the objective is
replaced by the coefficient of variation of $w$ and the exact
balancing constraints are relaxed to tolerance level). Since that
problem corresponds to use the unconventional link function $l(p) =
1/p$, \citet{zubizarreta2015stable} needed to include the additional
constraint that $w$ is nonnegative.


\subsection{Inexact balance, multivariate two-sample test and
  bias-variance tradeoff}
\label{sec:bias-vari-trad}

When the CBSR maximization problem is regularized as in
\eqref{eq:regularized-glm}, its dual objective functions in
\eqref{eq:proper-rule-max-ps-att-dual} and
\eqref{eq:proper-rule-max-ps-dual} remain unchanged, but the covariate
balancing constraints are no longer exact. Consider the regularized
GLM approach in \Cref{sec:regul-regr} with $J(p_{\theta}) =
\|\theta\|_a^a/a$ for some $a \ge 1$, the dual constraints are
given by
\begin{equation} \label{eq:kkt-regularized}
  \Big| \sum_{T_i = 1} w_{\hat{\theta}_{\lambda}}(X_i, T_i) \phi_k(X_i) -
  \sum_{T_i = 0} w_{\hat{\theta}_{\lambda}}(X_i, T_i) \phi_k(X_i)
  \Big|
  \le
  \lambda \cdot |(\hat{\theta}_{\lambda})_k|^{a-1}, k=1,\dots,m.
\end{equation}
The equality in \eqref{eq:kkt-regularized} holds if
$(\hat{\theta}_{\lambda})_k \ne 0$, which is generally true unless $a =
1$.

Following \Cref{sec:impl-covar-balance}, if we assume constant
treatment effect
$\mathrm{E}[Y(1)|X] \equiv \mathrm{E}[Y(0)|X] + \tau^{*}$ and the outcome
regression function is in the linear span of the regressors
$g_0(x) = g_{{\beta}}({x}) = \sum_{k=1}^m
\beta_k \phi_k({x})$, then the absolute bias of $\hat{\tau}^{*}_{\lambda}$ is
\begin{equation*}
  \Bigg| \sum_{k=1}^{\infty} \beta_k \cdot \Big[\sum_{T_i=1}
  w_i \phi_k(X_i) - \sum_{T_i=0} w_i \phi_k(X_i) \Big]
  \Bigg| \le \lambda \sum_{k=1}^{\infty} |\beta_k| \cdot
  |(\hat{\theta}_{\lambda})_k|^{a-1} \le \lambda \|\beta\|_a \|\hat{\theta}_{\lambda}\|_a^{a-1}.
\end{equation*}
The last inequality is due to H\"{o}lder's inequality and is tight.
In other words, the dual constraints imply that
\begin{equation} \label{eq:max-bias}
  \sup_{\|\beta\|_a \le 1} \big|\mathrm{bias}(\hat{\tau}^{*}_{\lambda}, g_{\beta})\big| = \lambda
  \|\hat{\theta}_{\lambda}\|_a^{a-1},~a \ge 1.
\end{equation}

The next proposition states that the right hand side of the last equation
is decreasing as the degree of regularization $\lambda$ becomes
smaller. This is consistent with the heuristic that the more we
regularize the propensity score model, the more bias our estimator is.

\begin{proposition} \label{prop:reguarlized-bias}
  Given a strictly proper scoring rule $S$ and a
  link function $l$ such that $S(l^{-1}(f),t)$ is strongly concave and second order
  differentiable in $f \in \mathbb{R}$ for $t = 0, 1$, let $\hat{\theta}_{\lambda}$ be
  the solution to \eqref{eq:regularized-glm} with $J(p_{\theta}) =
  \|\theta\|_a^a/a$ for a given $a \ge 1$. Then $\lambda
  \|\hat{\theta}_{\lambda}\|_a^{a-1}$ is a strictly increasing function
  of $\lambda > 0$.
\end{proposition}

The Lagrangian dual problems \eqref{eq:proper-rule-max-ps-att-dual}
and \eqref{eq:proper-rule-max-ps-dual} highlight the bias-variance trade-off when
using CBSR to estimate the propensity score. The dual objective
function measures the uniformity of $w$ (closely related to the variance of
$\hat{\tau}^{*}$) and the dual constraints bound the covariate imbalance
of $w$ (the minimax bias of $\hat{\tau}^{*}$ for $g_0(x) = g_{{\eta}}({x}) = \sum_{k=1}^m
\eta_k \phi_k({x})$ such that $\|\eta\|_a$ is less than a constant). The penalty
parameter $\lambda$ regulates this bias-variance trade-off.
When $\lambda \to 0$, the solution of
\eqref{eq:regularized-glm} converges to the weights $w$ that
minimizes the $a/(a-1)$-norm of covariate imbalance. The limit
of $r(\lambda)$ when $\lambda \to 0$ can be $0$ or some positive
value, depending on if the unregularized score maximization problem
\eqref{eq:opt-score-est} is feasible or not. When $\lambda \to
\infty$, the solution of \eqref{eq:regularized-glm} converges to
uniform weights (i.e.\ no adjustment at all) whose estimator $\hat{\tau}^{*}$ has
smallest variance.

A particularly interesting case is the lasso penalty $J(p_{\theta}) =
\|\theta\|_1$. By \eqref{eq:kkt-regularized}, the maximum covariate
imbalance is bounded by by $\lambda$. Therefore, the approximate
balancing weights proposed by \citet{zubizarreta2015stable} can be
viewed as putting weighted lasso penalty in propensity score
estimation. Bounding the maximum covariate imbalance can be useful
when the dimension of $X$ is high, see \citet{athey2016efficient}.

The RKHS regression in \Cref{sec:rkhs} is a generalization to the
regularized regression with potentially infinite-dimensional
predictors and weighted $l_2$-norm penalty. The maximum
bias under the sharp null is given by
\begin{equation} \label{eq:imbalance-g-regularized-max-rkhs}
  \sup_{\|g_0\|_{\mathcal{H}_K} \le 1}|\mathrm{bias}(\hat{\tau}^{*}_{\lambda},
  g_0)| \le \lambda \|\hat{f}_{\lambda}\|_{\mathcal{H}_K}.
\end{equation}

\begin{figure}[t]
  \centering
  \includegraphics[width = \textwidth]{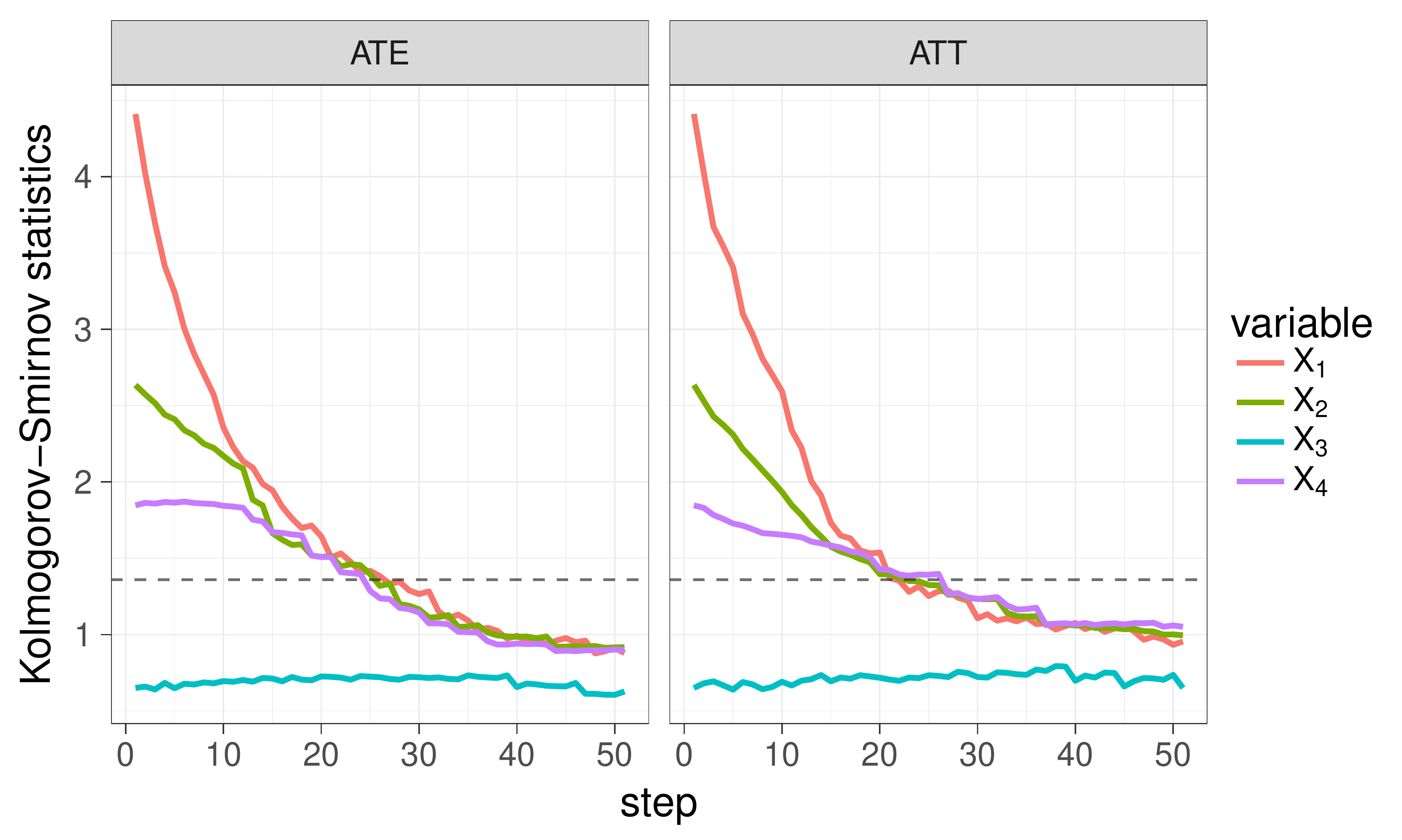}
  \caption{Boosting with $1$-level trees is reducing the maximum
    Kolmogorov-Smirnov statistics. Two estimands (ATT and ATE) and
    their corresponding CBSR are considered. Dashed line is the upper $0.05$
    quantile of the asymptotic null distribution of the K-S
    statistic.}
  \label{fig:boost-ks}
\end{figure}

The boosted trees in \Cref{sec:boosting} does not have a dual
problem since it is solved by a greedy algorithm. However, it shares a similar
interpretation with the lasso regularized GLM. With some algebra, the
gradient direction in \eqref{eq:g-grad} can be shown to be
\[
\hat{h} \propto \arg \max_{h \in \mathcal{F}_{k\textrm{-tree}}}
\Big[\sum_{T_i=1} w_i h(X_i) - \sum_{T_i=0} w_i h(X_i) \Big].
\]
That is, $\hat{h}$ is currently the most imbalanced $k$-tree. By
taking a small gradient step in the direction $\hat{h}$, it reduces
the covariate imbalance (bias) in this direction the fastest among all
$k$-trees. To see this, when $k=1$, maximum covariate imbalance among
$1$-trees is essentially the largest univariate Kolmogorov-Smirnov
statistics. We illustrate this interpretation of boosting using the
toy example in \Cref{sec:toy-example}. \Cref{fig:boost-ks} plots the
paths of Kolmogorov-Smirnov statistics as more trees are added to the
propensity score model (the step size is $\nu = 0.1$). The behavior is
similar to the lasso regularized CBSR-maximization
\eqref{eq:kkt-regularized} which reduces the largest univariate
imbalance (instead of the largest Kolmogorov-Smirnov statistic).

As a final remark, the left hand side of \eqref{eq:max-bias} or
\eqref{eq:imbalance-g-regularized-max-rkhs} indeed defines a distance
metric between two probability distributions (the empirical
distributions of the covariates over treatment and control). This
distance is called \emph{integral probability metric}
\citep{muller1997integral} and has received increasing attention
recently in the two-sample testing literature. In particular, a very
successful multivariate two-sample test \citep{gretton2012kernel} uses
the left hand side of \eqref{eq:imbalance-g-regularized-max-rkhs} as
its test statistic. Here we have given an alternative statistical
motivation of considering the integral probabability metric.

\subsection{A Bayesian interpretation}
\label{sec:bayes-interpr}

Besides the maximum bias interpretation
\eqref{eq:imbalance-g-regularized-max-rkhs}, the RKHS model in
\Cref{sec:rkhs} has another interesting Bayesian interpretation.
Suppose the regression function $g_0$ is also random and generated from
a Gaussian random field prior $g_0(\cdot) \sim \mathcal{G}(0,K)$ with mean
function $0$ and covariance function $K(\cdot,\cdot)$. Then the design
MSE of $\hat{\tau}^{*}$ (conditional on $\{X_i,T_i\}_{i=1}^n$)
under constant treatment effect is given by
\[
\begin{split}
&\mathrm{E}_g\Big[\sum_{T_i=1}
w_i g_0(X_i) - \sum_{T_i=0} w_i g_0(X_i) \Big]^2 +
\sum_{i=1}^n w_i^2 \mathrm{Var}(Y_i|X_i,T_i) \\ =& \tilde{w}^T K \tilde{w}
+ \sum_{i=1}^n w_i^2 \mathrm{Var}(Y_i|X_i,T_i),
\end{split}
\]
where $\tilde{w}_i = (2T_i - 1) w_i$, $i=1,\dotsc,n$ and (with some
abuse of notation) $K$ is the sample covariance matrix $K_{ij} =
K(X_i,X_j)$, $i,j=1,\dotsc,n$. This is directly tied to the dual
problem of CBSR maximization. For example, when the estimand is ATT
and the link is logistic, using the ``kernel trick'' $\hat{f}(X_i) = K
\hat{\gamma}$ described in \Cref{sec:rkhs} it is not difficult to show
the dual problem minimizes $\lambda \tilde{w}^T K \tilde{w} + \sum_{i=1}^n
w_i \log w_i$. Choosing different penalty parameters $\lambda$
essentially amounts to different prior beliefs about the conditional
variance $\mathrm{Var}(Y|X,T)$. We will explore this Bayesian
interpretation in a simulation example in
\Cref{sec:numerical-examples}.

In practice, optimally choosing the regularization parameter $\lambda$
is essentially difficult as it requires prior knowledge about
$\|g_0\|_{\mathcal{H}_K}$ and the conditional variance of $Y$
(essentially the signal-to-noise ratio). Such difficulty exists in all
previous approaches and we only attempt to provide a reasonable
solution here. Our experience with the
adaptive procedures in \Cref{sec:extensions} is
that once $\lambda$ is reasonably small, the further
reduction of maximum bias by decreasing $\lambda$ becomes
negligible in most cases. Our best recommendation is to plot the curve
of the maximum bias versus $\lambda$, and then the user should use her best
judgment based on prior knowledge about the outcome regression. To
mitigate the problem of choosing $\lambda$, next we describe how to make
valid statistical inference with an arbitrarily chosen
$\lambda$.

\subsection{Design-based finite-sample inference of WATE}
\label{sec:design-based-infer}

When the treatment effect is not homogeneous, the derivation above no
longer holds in general, although the bias-variance trade-off is still expected if the
effect is not too inhomogeneous. One exception is when the
estimand is ATT. In this case, if the weights are normalized so $w_i =
1$ if $T_i=1$, the finite sample bias of $\hat{\tau}^{*}$ is
\[
\begin{split}
&\Big[\sum_{T_i=1} w_i g_1(X_i) - \sum_{T_i=0} w_i g_0(X_i) \Big] -
\frac{1}{n_1} \Big[ \sum_{T_i=1} g_1(X_i) - g_0(X_i) \Big]\\ =&
\Big[\sum_{T_i=1} w_i g_0(X_i) - \sum_{T_i=0} w_i g_0(X_i) \Big].
\end{split}
\]
Therefore, the bias of $\hat{\tau}^{*}$ is only determined by how well $w$
balances $g_0$. This fact was noticed in
\citet{zhao2015,athey2016efficient,kallus2016generalized} and will be
used to construct honest confidence interval for the ATT.

To derive design-based inference of WATE, we assume strong ignorability (\Cref{assump:strong-ignor})
and $Y_i \sim \mathrm{N}(g_{T_i}(X_i), \sigma^2)$. The normality
assumption is not essential when sample size is large, but the
homoskedastic assumption is more difficult to relax. We assume the
treatment effect is constant if the estimand is not ATT. The only other
assumption we make is
\begin{assumption} \label{assump:outcome-reg}
  $g_0(x)$ is in a known RKHS $\mathcal{H}_K$.
\end{assumption}
Let the basis function of $\mathcal{H}_K$ be
$\{\phi_1(x),\phi_2(x),\dotsc,\}$ and $g_0(x) = \sum_{k=1}^{\infty}
\beta_k \phi_k(x)$. Suppose the propensity score is estimated by
the RKHS regression described in \Cref{sec:rkhs}.
Then by the decomposition \eqref{eq:bias-tau} and equation
\eqref{eq:imbalance-g-regularized-max-rkhs},
\[
|\hat{\tau}^{*} - \tau^{*}| \preceq \lambda \|g_0\|_{\mathcal{H}_K}
\|\hat{f}_{\lambda}\|_{\mathcal{H}_K} + \mathrm{N}\Big(0,
\sigma^2\sum_{i=1}^nw_i^2\Big),
\]
where $\preceq$ means stochastically smaller. Therefore, if we can
find an upper-$(\alpha/2)$ confidence limit for
$\|g_0\|_{\mathcal{H}_K}$ (denoted by
$\mathrm{CL}(\|g_0\|_{\mathcal{H}_K},1-\alpha/2)$) and a good estimate
of $\sigma$ (denoted by $\hat{\sigma}$), then a $(1 - \alpha)$-confidence interval of $\tau^*$ is given by
\begin{equation}
  \label{eq:tau-ci}
  \hat{\tau}^{*} \pm \big[\lambda \|\hat{f}_{\lambda}\|_{\mathcal{H}_K} \cdot
  \mathrm{CL}(\|g_0\|_{\mathcal{H}_K},1-\alpha/2) + \hat{\sigma}
  \|w\|_2 z_{1 - \alpha/2} \big],
\end{equation}
where $z_{1 - \alpha/2}$ is the upper-$(\alpha/2)$ quantile of the
standard normal distribution. This inferential method can be further
extended when an outcome regression adjustment is used (see
\Cref{sec:outcome-regression}) by replacing $g_0$ with $g_0 -
\hat{g}_0$. Notice that in this case $\hat{g}_0$ and
$\|\hat{g}_0\|_{\mathcal{H}_K}$ should be estimated using independent
sample in order to maintain validity of \eqref{eq:tau-ci}.

Note that our \Cref{assump:outcome-reg} also covers the setting where $X$
is high dimensional ($d \gg n$) and $g_0(x) = \sum_{k=1}^d \beta_k
X_k$. In this case, estimating $\|g_0\|_{\mathcal{H}_K} = \|\beta\|_2^2$ is of high
interest in genetic heritability and we shall use a recent proposal by
\citet{janson2016eigenprism} in our numeric example
below. Estimating $\|g_0\|_{\mathcal{H}_K}$ when $X$ is
low-dimensional can be done in a similar manner by weighting the
coefficients.

\citet{athey2016efficient} considered the inference of
ATT when $X$ is high dimensional, but a crucial assumption they
require is that $\beta$ is a very sparse vector so that $g_0$ can be
accurately estimated by lasso regression. In this case, $\lambda
\|\hat{g}_0 - g_0\|_{\mathcal{H}_K}
\|\hat{f}_{\lambda}\|_{\mathcal{H}_K}$ is negligible if $\lambda$ is
carefully chosen. Our confidence interval above does not require
the sparsity assumption since the procedure in
\citet{janson2016eigenprism} does not need sparsity. Balancing
functions in a kernel space is also considered in
\citet{hazlett2013balancing} and \citet{kallus2016generalized}, but
they did not consider the statistical inference of weighted
average treatment effects.


\section{Numerical Examples}
\label{sec:numerical-examples}


This Section provide two simulation examples to
demonstrate the flexibility of the proposed framework.

\subsection{Simulation: low-dimensional covariates}
\label{sec:low-dim}

To illustrate the bias-variance trade-off in selecting model space and
regularization parameter, in the following simulation we use a random
regression function instead of a manually selected regression function
to generate outcome observations. This is motivated by the Bayesian
interpretation in \Cref{sec:bias-vari-trad}. We believe this novel
simulation design also better reflects the philosophy of design-based
causal inference---the weights generated by the estimated propensity
score should be robust against any reasonable outcome regression
function.

In this simulation, we consider propensity score models fitted using
six kernels: the Gaussian kernel $k(x,x') = \exp(-\sigma\|x-x'\|^2)$
with $\sigma = 0.1$ or $1$, the Laplace kernel $k(x,x') = \exp(-\sigma\|x-x'\|)$
with $\sigma = 0.1$ or $1$, and the polynomial kernel $k(x,x') = (x^T
x' + 0.5)^d$ with $d = 1$ or $3$. Note that in Gaussian and Laplace
kernels, smaller $\sigma$ indicates more smoothness. The sample size
is $n = 1000$ and the covariates are generated by $X_i
\overset{\mathrm{i.i.d.}}{\sim} \mathrm{N}(0, I_5)$. The true
propensity score is a random function generated by
$\mathrm{logit}(P(T=1|X=x)) = f(x) \sim
\mathcal{G}(0,K(\cdot,\cdot))$ where the covariance function is
either the polynomial kernel with degree $1$ or the Gaussian kernel
with $\sigma = 0.1$. Potential outcomes are generated from
the sharp null model $Y_i(0) = Y_i(1) = g_0(X_i) + \epsilon_i$ where
$g_0(X_i)$ is a random function generated by the same Gaussian process
with any of the six considered kernels. Note that for Gaussian and
Laplace kernels, smaller $\sigma$ indicates smoother random
functions. For the polynomial kernels, a randomly generated function
is just a linear or cubic function with random coefficients.

For choosing the regularization parameter $\lambda$, we stop the
RKHS regression described in \Cref{sec:rkhs} when the coefficient of
variation of $w$ is less than $0.5$ (``stop early'' in the table) or
less than $1.2$ (``stop late'' in the table). We consider two
estimators, the weighted difference estimator with no outcome
adjustment (IPW in the table) or with adjustment by outcome regression that is
fitted by kernel least squares (same kernel in estimating the
propensity score) and tuned by cross-validation (AIPW in the table).

\Cref{tab:kernel} reports the average absolute bias over 100
simulations of the two estimators under different simulation settings
and propensity score models. Some observations from this table:
\begin{enumerate}
\item Under all settings, the lowest bias is always achieved when the
  fitting kernel is the same kernel that generates the outcome
  regression function $g_0$. This is expected from our Bayesian
  interpretation in \Cref{sec:bayes-interpr}.
\item Outcome adjustment is very helpful when the propensity score
  estimation is stopped early (so the covariates are less
  balanced), but makes little improvement when the propensity score
  estimation is stopped late.
\item There is no uniformly best performing kernel in fitting the
  propensity score. In particular, polynomial kernels perform poorly
  when $g_0$ is not a polynomial. Laplace kernel with $\sigma = 0.1$
  performs relatively well in most of the simulation settings.
\item Surprisingly, the kernel used to generate $f$ (logit of the true propensity
  score) does not alter the qualitative conclusions above. Even if the
  ``correct'' kernel is used to fit the propensity score model, there is
  no guarantee that this better reduces the average bias than other
  ``incorrect'' kernels. For example, when $f$ is simulated from
  poly($d=1$), i.e.\ $f$ is a random linear function, using the linear
  propensity score model performs poorly unless $g_0$ is also a linear
  function.
\end{enumerate}
In \Cref{sec:conclusions}, we discuss the practical implications of
this simulation result in more detail.

\begin{table}
  \centering
  \caption{Simulation: low-dimensional covariates. Reported numbers
    are average absolute bias over $100$ simulations.}
  \label{tab:kernel}
  \renewcommand{\arraystretch}{1.1}
\begin{tabular}{ll|cccc|cccc}
\hline
\multicolumn{2}{c|}{$f$} & \multicolumn{4}{c|}{poly($d=1$)} & \multicolumn{4}{c}{gau($\sigma=0.1$)} \\
\multicolumn{2}{c|}{$\lambda$} & \multicolumn{2}{c}{stop early} &
\multicolumn{2}{c|}{stop late} &
\multicolumn{2}{c}{stop early} & \multicolumn{2}{c}{stop late} \\
\hline
$g_0$ & fitting kernel & IPW & AIPW & IPW & AIPW & IPW & AIPW & IPW & \multicolumn{1}{c}{AIPW} \\
\hline
lap($\sigma=0.1$) & lap($\sigma=0.1$)  & $0.58$ & $0.32$ & $0.16$ & $0.15$ & $0.48$ & $0.30$ & $0.20$ & $0.20$ \\
 & lap($\sigma=1$)  & $0.59$ & $0.39$ & $0.24$ & $0.24$ & $0.50$ & $0.30$ & $0.22$ & $0.22$ \\
 & poly($d=1$)  & $0.60$ & $0.41$ & $0.21$ & $0.22$ & $0.50$ & $0.33$ & $0.29$ & $0.29$ \\
 & poly($d=3$)  & $0.63$ & $0.40$ & $0.25$ & $0.24$ & $0.55$ & $0.37$ & $0.25$ & $0.25$ \\
 & gau($\sigma=0.1$)  & $0.59$ & $0.34$ & $0.20$ & $0.21$ & $0.48$ & $0.31$ & $0.24$ & $0.24$ \\
 & gau($\sigma=1$)  & $0.62$ & $0.50$ & $0.35$ & $0.35$ & $0.58$ & $0.41$ & $0.29$ & $0.30$ \\
lap($\sigma=1$) & lap($\sigma=0.1$)  & $0.78$ & $0.49$ & $0.46$ & $0.46$ & $0.65$ & $0.39$ & $0.43$ & $0.43$ \\
 & lap($\sigma=1$)  & $0.79$ & $0.48$ & $0.43$ & $0.43$ & $0.64$ & $0.40$ & $0.38$ & $0.38$ \\
 & poly($d=1$)  & $0.80$ & $0.61$ & $0.52$ & $0.52$ & $0.69$ & $0.51$ & $0.50$ & $0.50$ \\
 & poly($d=3$)  & $0.84$ & $0.58$ & $0.58$ & $0.58$ & $0.72$ & $0.46$ & $0.51$ & $0.51$ \\
 & gau($\sigma=0.1$)  & $0.78$ & $0.50$ & $0.56$ & $0.56$ & $0.65$ & $0.38$ & $0.52$ & $0.52$ \\
 & gau($\sigma=1$)  & $0.80$ & $0.56$ & $0.50$ & $0.51$ & $0.66$ & $0.50$ & $0.41$ & $0.42$ \\
poly($d=1$) & lap($\sigma=0.1$)  & $1.42$ & $0.34$ & $0.10$ & $0.08$ & $1.21$ & $0.22$ & $0.07$ & $0.06$ \\
 & lap($\sigma=1$)  & $1.47$ & $0.69$ & $0.46$ & $0.46$ & $1.33$ & $0.47$ & $0.35$ & $0.34$ \\
 & poly($d=1$)  & $1.39$ & $0.18$ & $0.04$ & $0.00$ & $1.11$ & $0.14$ & $0.00$ & $0.00$ \\
 & poly($d=3$)  & $1.46$ & $0.47$ & $0.10$ & $0.05$ & $1.29$ & $0.36$ & $0.03$ & $0.03$ \\
 & gau($\sigma=0.1$)  & $1.44$ & $0.40$ & $0.14$ & $0.10$ & $1.22$ & $0.27$ & $0.07$ & $0.07$ \\
 & gau($\sigma=1$)  & $1.60$ & $1.17$ & $0.82$ & $0.83$ & $1.50$ & $0.89$ & $0.66$ & $0.64$ \\
poly($d=3$) & lap($\sigma=0.1$)  & $1.16$ & $0.48$ & $0.31$ & $0.30$ & $0.94$ & $0.45$ & $0.32$ & $0.31$ \\
 & lap($\sigma=1$)  & $1.25$ & $0.66$ & $0.53$ & $0.53$ & $0.98$ & $0.54$ & $0.48$ & $0.48$ \\
 & poly($d=1$)  & $1.17$ & $0.64$ & $0.57$ & $0.56$ & $0.98$ & $0.74$ & $0.73$ & $0.73$ \\
 & poly($d=3$)  & $1.15$ & $0.34$ & $0.17$ & $0.09$ & $0.92$ & $0.34$ & $0.06$ & $0.03$ \\
 & gau($\sigma=0.1$)  & $1.18$ & $0.44$ & $0.26$ & $0.22$ & $0.95$ & $0.42$ & $0.23$ & $0.21$ \\
 & gau($\sigma=1$)  & $1.32$ & $1.02$ & $0.79$ & $0.79$ & $1.04$ & $0.76$ & $0.63$ & $0.63$ \\
gau($\sigma=0.1$) & lap($\sigma=0.1$)  & $0.89$ & $0.34$ & $0.16$ & $0.15$ & $0.81$ & $0.35$ & $0.12$ & $0.12$ \\
 & lap($\sigma=1$)  & $0.95$ & $0.40$ & $0.27$ & $0.27$ & $0.81$ & $0.39$ & $0.19$ & $0.19$ \\
 & poly($d=1$)  & $0.92$ & $0.46$ & $0.37$ & $0.37$ & $0.85$ & $0.58$ & $0.51$ & $0.51$ \\
 & poly($d=3$)  & $0.95$ & $0.42$ & $0.22$ & $0.20$ & $0.83$ & $0.41$ & $0.18$ & $0.18$ \\
 & gau($\sigma=0.1$)  & $0.90$ & $0.36$ & $0.12$ & $0.11$ & $0.80$ & $0.31$ & $0.07$ & $0.08$ \\
 & gau($\sigma=1$)  & $0.99$ & $0.62$ & $0.45$ & $0.45$ & $0.84$ & $0.57$ & $0.34$ & $0.33$ \\
gau($\sigma=1$) & lap($\sigma=0.1$)  & $0.59$ & $0.60$ & $0.48$ & $0.49$ & $0.64$ & $0.57$ & $0.44$ & $0.46$ \\
 & lap($\sigma=1$)  & $0.57$ & $0.49$ & $0.42$ & $0.42$ & $0.62$ & $0.53$ & $0.45$ & $0.46$ \\
 & poly($d=1$)  & $0.59$ & $0.58$ & $0.58$ & $0.58$ & $0.64$ & $0.64$ & $0.62$ & $0.62$ \\
 & poly($d=3$)  & $0.62$ & $0.56$ & $0.65$ & $0.66$ & $0.65$ & $0.62$ & $0.55$ & $0.57$ \\
 & gau($\sigma=0.1$)  & $0.58$ & $0.61$ & $0.60$ & $0.63$ & $0.62$ & $0.63$ & $0.64$ & $0.67$ \\
 & gau($\sigma=1$)  & $0.55$ & $0.46$ & $0.41$ & $0.41$ & $0.60$ & $0.53$ & $0.42$ & $0.43$ \\
\hline
\end{tabular}
\end{table}

\subsection{Simulation: high-dimensional covariates}
\label{sec:high-dim}

In our second example, we consider the case that the covariates $X$
are high dimensional. In this simulation, the sample size $n = 1000$
and $X_i \in \mathbb{R}^{100} \overset{\mathrm{i.i.d.}}{\sim}
\mathrm{N}(0, \Sigma)$ where $\Sigma_{ij} = 0.5^{|i-j|}$. The true
propensity score is $\mathrm{logit}(P(T_i=1|X_i)) = \rho X_i^T \theta$
where $\rho = 1~\mathrm{or}~2$, $\theta$ is a $100$-dimensional vector
whose first $s_t$ entries are $1/\sqrt{s_t}$ and the rest are zero, and
$s_t = 5~\mathrm{or}~100$. The potential outcomes are generated from
the sharp null model $Y_i(0) = Y_i(1) = X_i^T \beta + \epsilon_i$,
where the first $s_y$ entries of $\beta$ are $1/\sqrt{s_y}$ and the
rest are zero, $s_y = 5,~20,~\mathrm{or}~100$, and $\epsilon_i$ is an
independent Gaussian noise with standard deviation $\sigma=5$.

In this simulation, the propensity score model is fitted by maximizing the
CBSR corresponding
to ATT ($\alpha = 0$, $\beta = -1$) with ridge penalty. The
regularization parameter $\lambda$ is chosen so that the coefficient
of variation of the weights is just below $1$. Three estimators are
considered: the weighted difference estimator with no outcome
adjustment (IPW), outcome adjustment fitted by the lasso (AIPW-L), and
outcome adjustment fitted by the ridge regression (AIPW-R). The
outcome regressions, either fitted by the lasso or the ridge penalty,
are tuned by cross-validation.

Averaging over $1000$ simulations, we report in \Cref{tab:linear} the
root-mean-square error of the estimators (RMSE), the absolute bias
(Bias), the estimated
maximum bias as described in \Cref{sec:design-based-infer} which uses
the EigenPrism method of \citet{janson2016eigenprism} to estimate
$\|\beta\|_2$ (Max Bias), coverage of the 95\%-confidence interval ignoring
covariate imbalance as in \citet{athey2016efficient} (CI), coverage of
the honest 95\%-confidence interval \eqref{eq:tau-ci} (Honest CI), and the
ratio of the length between the two confidence intervals (CI Ratio).

Different from the simulation with low-dimensional covariates in
\Cref{sec:low-dim}, outcome regression adjustment improves estimation
accuracy quite significantly. As expected, when $\theta$ is dense
ridge outcome regression performs better and when $\theta$ is sparse
lasso outcome regression performs better. In many settings, the actual
bias is a substantial portion of the estimated maximum bias. Ignoring
this bias in the construction of confidence interval can lead to
serious under-coverage of the causal parameter, as indicated by the CI
column in \Cref{tab:linear}. Note that the sparsity assumption in
\citet{athey2016efficient} requires $s_y \ll \sqrt{n}/\log(d) \approx
6.9$, so the lack of coverage does not violate the theoretical results
in \citet{athey2016efficient} as the smallest $s_y$ in this simulation
is $5$. Using the honest confidence interval
derived in \Cref{sec:design-based-infer} ensures the desired coverage,
although the confidence interval is a few times wider and quite
conservative as expected.

\begin{table}
  \centering
  \caption{Simulation: high-dimensional covariates. Reported values
    are the average RMSE, average absolute bias, average estimated
    maximum bias, coverage of the confidence interval ignoring bias due to
  inexact balance, coverage of the honest confidence interval proposed
in \Cref{sec:design-based-infer}, and the average ratio between the
two confidence intervals over $1000$ simulations.}
  \label{tab:linear}
  \renewcommand{\arraystretch}{1.1}
  \begin{tabular}{llll|c|c|c|c|c|c}
    \hline
    $s_y$ & $\rho$ & $s_t$ & Method & RMSE & Bias & Max Bias & CI &
    Honest CI & CI Ratio \\
\hline
100 & 1 & 100 & IPW  & $1.41$ & $1.35$ & $1.47$ & $0.12$ & $0.99$ & $6.00$ \\
 &  &  & AIPW-L  & $1.30$ & $1.21$ & $1.55$ & $0.23$ & $0.99$ & $6.08$ \\
 &  &  & AIPW-R  & $1.22$ & $1.11$ & $1.44$ & $0.30$ & $0.99$ & $5.96$ \\
 &  & 5 & IPW  & $0.51$ & $0.28$ & $1.41$ & $0.88$ & $1.00$ & $5.84$ \\
 &  &  & AIPW-L  & $0.52$ & $0.25$ & $1.58$ & $0.87$ & $0.98$ & $5.95$ \\
 &  &  & AIPW-R  & $0.51$ & $0.21$ & $1.32$ & $0.87$ & $0.98$ & $5.66$ \\
 & 2 & 100 & IPW  & $0.91$ & $0.78$ & $1.02$ & $0.49$ & $1.00$ & $4.40$ \\
 &  &  & AIPW-L  & $0.83$ & $0.66$ & $1.10$ & $0.57$ & $1.00$ & $4.46$ \\
 &  &  & AIPW-R  & $0.76$ & $0.57$ & $0.98$ & $0.65$ & $1.00$ & $4.34$ \\
 &  & 5 & IPW  & $0.42$ & $0.10$ & $0.59$ & $0.96$ & $0.99$ & $2.96$ \\
 &  &  & AIPW-L  & $0.43$ & $0.09$ & $0.67$ & $0.95$ & $0.98$ & $3.00$ \\
 &  &  & AIPW-R  & $0.43$ & $0.07$ & $0.55$ & $0.96$ & $0.98$ & $2.88$ \\
20 & 1 & 100 & IPW  & $0.71$ & $0.60$ & $1.53$ & $0.74$ & $1.00$ & $6.12$ \\
 &  &  & AIPW-L  & $0.62$ & $0.43$ & $1.61$ & $0.82$ & $1.00$ & $6.13$ \\
 &  &  & AIPW-R  & $0.63$ & $0.44$ & $1.44$ & $0.82$ & $1.00$ & $5.95$ \\
 &  & 5 & IPW  & $0.78$ & $0.64$ & $1.42$ & $0.62$ & $1.00$ & $5.86$ \\
 &  &  & AIPW-L  & $0.71$ & $0.51$ & $1.47$ & $0.69$ & $0.99$ & $5.85$ \\
 &  &  & AIPW-R  & $0.66$ & $0.45$ & $1.36$ & $0.75$ & $0.99$ & $5.71$ \\
 & 2 & 100 & IPW  & $0.53$ & $0.34$ & $0.96$ & $0.87$ & $1.00$ & $4.38$ \\
 &  &  & AIPW-L  & $0.49$ & $0.24$ & $1.01$ & $0.90$ & $0.98$ & $4.37$ \\
 &  &  & AIPW-R  & $0.49$ & $0.24$ & $0.91$ & $0.89$ & $0.99$ & $4.26$ \\
 &  & 5 & IPW  & $0.48$ & $0.22$ & $0.56$ & $0.93$ & $0.99$ & $2.94$ \\
 &  &  & AIPW-L  & $0.47$ & $0.16$ & $0.59$ & $0.94$ & $0.99$ & $2.93$ \\
 &  &  & AIPW-R  & $0.46$ & $0.15$ & $0.52$ & $0.94$ & $0.99$ & $2.88$ \\
5 & 1 & 100 & IPW  & $0.51$ & $0.29$ & $1.59$ & $0.90$ & $0.99$ & $6.17$ \\
 &  &  & AIPW-L  & $0.49$ & $0.18$ & $1.55$ & $0.92$ & $0.98$ & $6.02$ \\
 &  &  & AIPW-R  & $0.52$ & $0.23$ & $1.54$ & $0.89$ & $0.98$ & $6.04$ \\
 &  & 5 & IPW  & $1.24$ & $1.19$ & $1.27$ & $0.17$ & $0.99$ & $5.73$ \\
 &  &  & AIPW-L  & $1.05$ & $0.91$ & $1.29$ & $0.39$ & $0.99$ & $5.75$ \\
 &  &  & AIPW-R  & $1.09$ & $0.99$ & $1.26$ & $0.33$ & $1.00$ & $5.72$ \\
 & 2 & 100 & IPW  & $0.43$ & $0.17$ & $0.98$ & $0.94$ & $0.99$ & $4.44$ \\
 &  &  & AIPW-L  & $0.42$ & $0.10$ & $0.93$ & $0.94$ & $0.99$ & $4.34$ \\
 &  &  & AIPW-R  & $0.43$ & $0.13$ & $0.92$ & $0.93$ & $0.99$ & $4.35$ \\
 &  & 5 & IPW  & $0.61$ & $0.43$ & $0.54$ & $0.82$ & $1.00$ & $2.93$ \\
 &  &  & AIPW-L  & $0.55$ & $0.29$ & $0.53$ & $0.88$ & $0.99$ & $2.91$ \\
 &  &  & AIPW-R  & $0.56$ & $0.33$ & $0.53$ & $0.87$ & $1.00$ & $2.91$ \\
\hline
  \end{tabular}
\end{table}


\section{Discussion}
\label{sec:conclusions}

We have proposed a general method of obtaining covariate balancing
propensity score which unifies many previous approaches. Our proposal
is conceptually simple: the investigator just needs to tailor the
loss function according to the link function and estimand. This offers
great flexibility in incorporating adaptive strategies developed in
statistical learning. We have given a through discourse on the dual
interpretation of minimizing the tailored loss function, especially
how regularization is linked to the bias-variance trade-off in
estimating the weighted average treatment effects. We provide honest
 inference that account for the bias incurred by inexact balance.

Throughout the paper we have taken an outright design perspective: without
looking at the outcome data, the investigator tries to balance
pre-treatment covariates as well as possible to mimic a randomized
experiment, echoing the recommendations by
\citet{rubin2008objective,rubin2009should}. This allows us to give an
interesting Bayesian
interpretation of covariate balance: when checking covariate imbalance
and deciding which
propensity score model is ``acceptable'', the investigator is
implicitly assuming a prior on the unknown outcome
regression function.

The conclusions from the simulation examples in
\Cref{sec:numerical-examples} are quite intricate and it stresses an
essential difficulty in observational studies: no design is uniformly
the best (unless the observations can be exactly matched). The
investigator must be aware that weaker modeling assumption ($g_0$ is in
a larger function space) leads to wider confidence interval of the
causal effect, and
any statistical inference is relative to this assumption. In practice,
when the covariate dimension is low we recommend using a
universal kernel (such as Gaussian or Laplace kernel, see e.g.\
\citet{gretton2012kernel}) to estimate the propensity score, so the
model space is dense in the space of continuous functions. We encourage
the user to try different kernels (e.g. Laplace kernel with different
$\sigma$) and report how the confidence interval changes with different
modeling assumption. This shows how sensitive the statistical
inference is to the modeling assumption. When the covariate dimension
is high and a linear outcome model is assumed, our simulation results
suggest that the user should be cautious about any further sparsity
assumption that eliminates the bias due to inexact balance. Honest
confidence interval can be constructed without the sparsity
assumption, but the interval can also be too wide if the covariate
dimension is much larger than the sample size.



\appendix

\section{Technical proofs}
\label{sec:theoretical-proofs}

\subsection{Proof of \Cref{prop:beta-concave}}
\label{sec:proof-crefpr-conc}

The same result can be found in \citet[Section
15]{buja2005loss}. For completeness we give a direct
proof here. Denote $p = l^{-1}(f) \in (0,1)$. Since $v =
v_{\alpha,\beta}$, we have
\[
\frac{G''(p)}{l'(p)} = p^{\alpha} (1-p)^{\beta}
\]
Therefore, by the chain rule and the inverse function theorem,
\begin{equation*} \label{eq:S-1st-derivative}
  \begin{split}
    \frac{\mathrm{d}}{\mathrm{d}f} S(l^{-1}(f),1) &= (1 - p) G''(p) (l^{-1})'(f)  = p^{\alpha}
    (1-p)^{\beta+1},\\
    \frac{\mathrm{d}}{\mathrm{d}f}  S(l^{-1}(f),0) &= - p G''(p) (l^{-1})'(f) = -
    p^{\alpha+1} (1-p)^{\beta},~\mathrm{and}
  \end{split}
\end{equation*}
\begin{equation*} \label{eq:S-2nd-derivative}
  \begin{split}
    \frac{\mathrm{d}^2}{\mathrm{d}f^2} S(l^{-1}(f),1) &= \alpha p^{\alpha} (1-p)^{\beta + 2} - (\beta + 1)
    p^{\alpha + 1} (1-p)^{\beta+1}, \\
    \frac{\mathrm{d}^2}{\mathrm{d}f^2} S(l^{-1}(f),0) &= - (\alpha + 1)
    p^{\alpha + 1} (1-p)^{\beta+1} + \beta p^{\alpha+2} (1-p)^{\beta}.
  \end{split}
\end{equation*}
The conclusions immediate follow by letting
the second order derivatives be less than or equal to $0$.

\subsection{Proof of \Cref{thm:beta-efficient}}
\label{sec:proof-crefthm:b-effi}

First we list the technical assumptions in \citet{hirano2003}:
\begin{assumption} (Distribution of ${X}$) \label{assump:X}
  The support of ${X}$ is a Cartesian product of compact
  intervals. The density of ${X}$ is bounded, and bounded away from
  $0$.
\end{assumption}

\begin{assumption} (Distribution of $Y(0),~Y(1)$) \label{assump:Y}
  The second moments of $Y(0)$ and $Y(1)$ exist and $g({X},0) =
  \mathrm{E}[Y(0)|{X}]$ and $g({X},1) = \mathrm{E}[Y(1)|{X}]$
  are continuously differentiable.
\end{assumption}

\begin{assumption} (Propensity score) \label{assump:ps}
  The propensity score $p({X}) = \mathrm{P}(T=1|{X})$ is
  continuously differentiable of order $s \ge 7d$ where $d$ is the
  dimension of ${X}$, and $p({x})$ is bounded away from $0$ and
  $1$.
\end{assumption}

\begin{assumption} (Sieve estimation) \label{assump:sieve}
  The nonparametric sieve logistic regression uses a power series with
  $m = n^{\nu}$ for some $1/(4(s/d-1)) < \nu < 1/9$.
\end{assumption}

The proof is a simple modification of the proof in
\citet{hirano2003}. In fact,
\citet{hirano2003} only proved the convergence of the estimated
propensity score up to certain order. This essentially suggests that the
semiparametric efficiency of $\hat{\tau}$ does not heavily depend on the accuracy of the sieve logistic regression.

To be more specific, only
three properties of the maximum likelihood rule $S = S_{0,0}$ are used in \citet[Lemmas 1,2]{hirano2003}:
\begin{enumerate}[label=\arabic*.]
\item $\tilde{{\theta}} = \arg \max_{{\theta}}
  S(p_{{\theta}},p_{\tilde{{\theta}}})$ (line 5, page 19), this is
  exactly the definition of a strictly proper scoring rule \eqref{eq:proper-rule};
\item The Fisher information matrix
  \[
  \frac{\partial^2}{\partial {\theta} \partial {\theta}^T}
  S(p_{{{\theta}}},p_{\tilde{{\theta}}}) = \mathrm{E}_{\tilde{{\theta}}}
  \left\{ \left[ \frac{\mathrm{d}^2}{\mathrm{d}f^2} S(l^{-1}(f),
      T)\Bigr|_{f={\phi}({X})^T {{\theta}}} \right] {\phi}({X}) {\phi}({X})^T \right\}
  \]
  has all eigenvalues uniformly bounded away from $0$ for all
  ${\theta}$ and $\tilde{{\theta}}$ in a compact set in $\mathbb{R}^m$, where the expectation on the right hand side is taken
  over ${X}$ and $T|{X} \sim p_{\tilde{{\theta}}}$.
\item As $m \to \infty$, with probability tending to $1$ the observed Fisher
  information matrix
  \[
  \frac{\partial^2}{\partial {\theta} \partial {\theta}^T}
  \frac{1}{n} \sum_{i=1}^n S(p_{{{\theta}}}({X}_i),T_i) =
  \frac{1}{n} \sum_{i=1}^n \left[\frac{\mathrm{d}^2}{\mathrm{d}f^2} S(l^{-1}(f),
    T_i)\Bigr|_{f={\phi}({X}_i)^T {{\theta}}} \right] {\phi}({X}_i) {\phi}({X}_i)^T
  \]
  has all eigenvalues uniformly bounded away from $0$ for all
  ${{\theta}}$ in a compact set of $\mathbb{R}^m$ (line 7--9, page 21).
\end{enumerate}
Because the approximating functions
${\phi}$ are obtained through orthogonalizing the power series,
we have $\mathrm{E}[{\phi}({X}) {\phi}({X})^T] =
{I}_m$ and one can show its finite sample version has eigenvalues bounded away from $0$ with probability going to $1$ as $n \to
\infty$. Therefore a
sufficient condition for the second and
third properties above is that $S(l^{-1}(f),t)$ is strongly concave for
$t=0,1$. In \Cref{prop:beta-concave} we have already proven the
strong concavity for all $-1 \le \alpha,\beta \ge 1$ except for
$\alpha = -1, \beta = 0$ and $\alpha = 0, \beta = -1$. In these
two boundary cases, among $S(l^{-1}(f),0)$ and $S(l^{-1}(f),1)$ one score
function is strongly concave and the other score function is linear
in $f$. One can still prove the second and third
properties by using \Cref{assump:ps} that the propensity score is
bounded away from $0$ and $1$.

\subsection{Proof of \Cref{prop:reguarlized-bias}}
\label{sec:proof-crefpr-bias}

The conclusion is trivial for $a = 1$. Denote
\[
h(f,t) = \frac{\mathrm{d}}{\mathrm{d}f}
S(l^{-1}(f),t)~\mathrm{and}~h'(f,t) = \frac{\mathrm{d}}{\mathrm{d}f} h(f,t),~t=0,1.
\]
Because $S(l^{-1}(f),t)$ is concave in $f$, we have $h'(f,t) < 0$ for all
$f$. The first-order optimality condition of
\eqref{eq:regularized-glm} is given by
\begin{equation*} 
  \frac{1}{n} \sum_{i=1}^n h(\hat{\theta}_{\lambda}^T{\phi}({X}_i),T_i)
  {\phi}_k({X}_i) + \lambda |(\hat{\theta}_{\lambda})_k|^{a-1} \mathrm{sign}((\hat{\theta}_{\lambda})_k) = 0,~k=1,\dotsc,m.
\end{equation*}
Let $\nabla \hat{\theta}_{\lambda}$ be the gradient of
$\hat{\theta}_{\lambda}$ with respect to $\lambda$. By taking
derivative of the identity above, we get
\[
\left[\frac{1}{n} \sum_{i=1}^n h'(\hat{\theta}_{\lambda}^T{\phi}_i,T_i)
  \phi_i \phi_i^T + \lambda (a-1)
  \mathrm{diag}(|\hat{\theta}_{\lambda}|^{a-2})\right] \nabla
\hat{\theta}_{\lambda} = - |\hat{\theta}_{\lambda}|^{a-1} \mathrm{sign}(\hat{\theta}_{\lambda}),
\]
where we used the abbreviation $\phi_i = \phi(X_i)$ and $\theta^a =
(\theta_1^a,\dotsc,\theta_m^a)$. For brevity, let's denote
\[
H = \frac{1}{n} \sum_{i=1}^n h'(\hat{\theta}_{\lambda}^T{\phi}_i,T_i)
\phi_i \phi_i^T \prec 0~\mathrm{and}~G = \lambda (a-1)
\mathrm{diag}(|\hat{\theta}_{\lambda}|^{a-2}).
\]

For $a > 1$, the result is proven by showing the derivative of $\lambda
\|\hat{\theta}_{\lambda}\|_a^{a-1}$ is greater than $0$.
\[
\begin{split}
  \frac{\mathrm{d}}{\mathrm{d} \lambda} \left( \lambda
    \|\hat{\theta}_{\lambda}\|_a^{a-1} \right)
  &= \|\hat{\theta}_{\lambda}\|_a^{a-1} + \lambda
  \frac{\mathrm{d}}{\mathrm{d} \lambda} \Bigg[ \sum_{j=1}^m
  \left|(\hat{\theta}_{\lambda})_k\right|^a \Bigg]^{(a-1)/a} \\
  &= \|\hat{\theta}_{\lambda}\|_a^{a-1} + \lambda (a-1)
  \|\hat{\theta}_{\lambda}\|_a^{-1} \sum_{j=1}^m
  \left|(\hat{\theta}_{\lambda})_k\right|^{a-1} (\nabla
  \hat{\theta}_{\lambda})_k \, \mathrm{sign}((\hat{\theta}_{\lambda})_k)
  \\
  &= \|\hat{\theta}_{\lambda}\|_a^{a-1} - \lambda (a-1)
  \|\hat{\theta}_{\lambda}\|_a^{-1} (|\hat{\theta}_{\lambda}|^{a-1})^T
  (H + G)^{-1} |\hat{\theta}_{\lambda}|^{a-1} \\
  &> \|\hat{\theta}_{\lambda}\|_a^{a-1} - \lambda (a-1)
  \|\hat{\theta}_{\lambda}\|_a^{-1} (|\hat{\theta}_{\lambda}|^{a-1})^T
  G^{-1} |\hat{\theta}_{\lambda}|^{a-1} \\
  &= 0.
\end{split}
\]


\section{A closer look at the Beta family}
\label{sec:closer-look-at}

\begin{figure}[p]
  \centering
  \begin{subfigure}[t]{0.8\textwidth}
    \includegraphics[width = \textwidth]{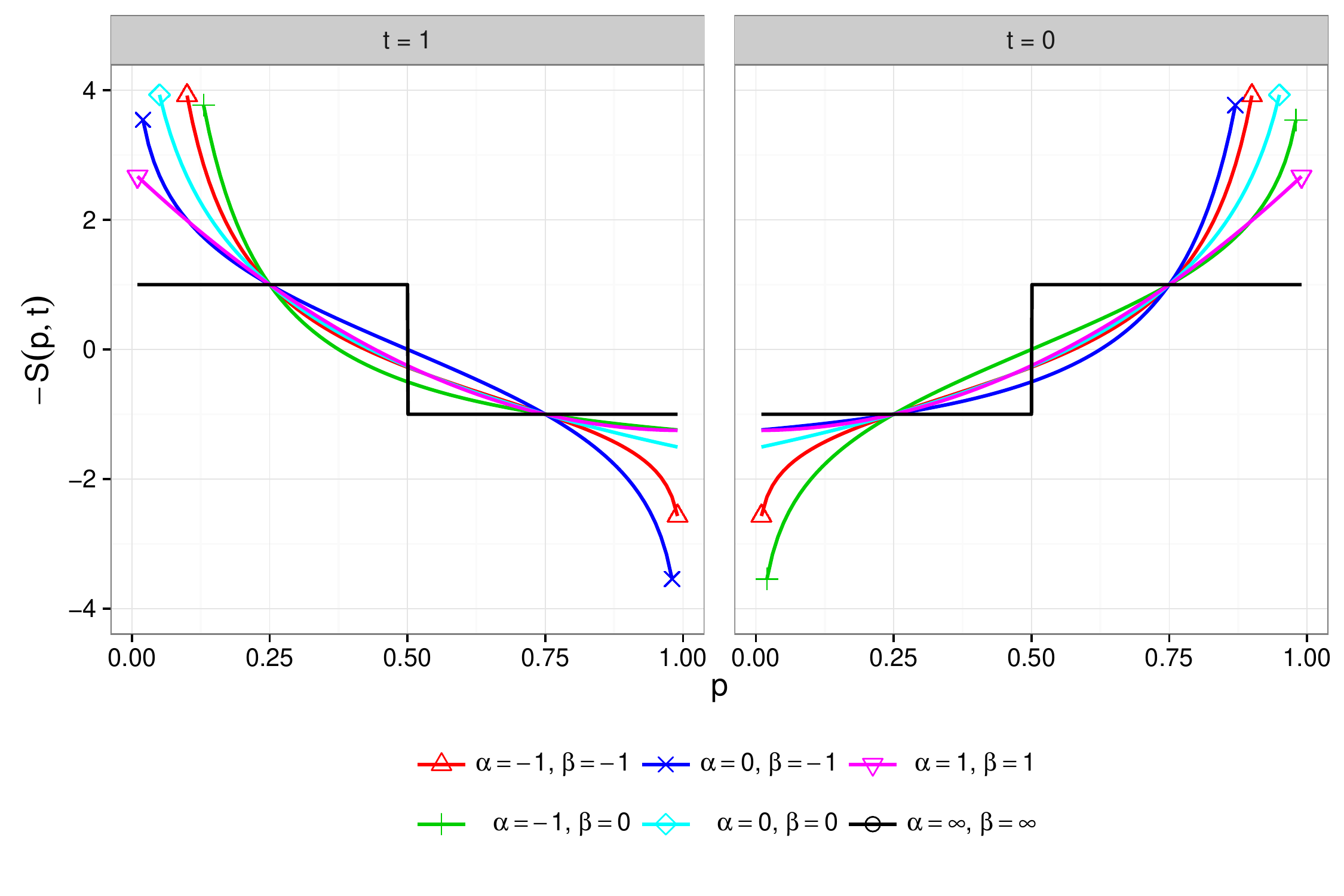}
    \caption{Loss functions $- S_{\alpha,\beta}(p, t)$ for $t = 0, 1$.}
    \label{fig:S-plot}
  \end{subfigure}

  \begin{subfigure}[t]{0.8\textwidth}
    \includegraphics[width = \textwidth]{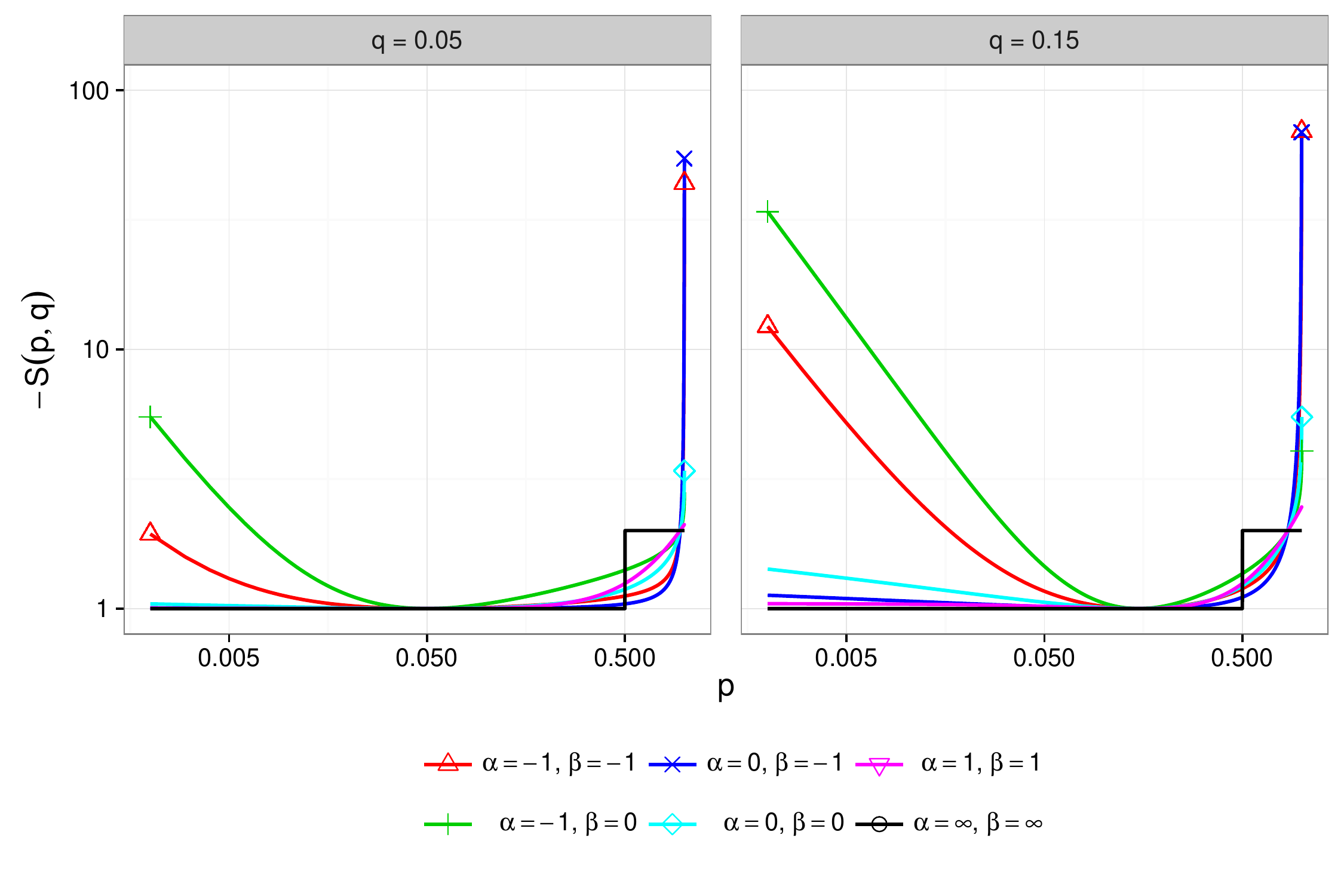}
    \caption{Loss functions $- S_{\alpha,\beta}(p, q)$ for $q =
      0.05$ and $0.15$.}
    \label{fig:S-pq-plot}
  \end{subfigure}

  \caption{Graphical illustration of the Beta-family of scoring rules defined in \eqref{eq:beta-family}.}
  \label{fig:S}

\end{figure}

\Cref{fig:S} plots the scoring rules
$S_{\alpha,\beta}$ for some combinations of $\alpha$ and $\beta$. The
top panels show the score function $S(p,0)$ and $S(p,1)$ for $0 <
p < 1$, which are normalized so that $S(1/4,1) = S(3/4,0) = -1$ and
$S(1/4,0) = S(3/4,1) = 1$.
By a change of variable, one can
show $S_{\alpha,\beta}(p,1) = S_{\beta,\alpha}(1 - p, 0)$. This is
the reason that the two subplots in \Cref{fig:S-plot} are essentially
reflections of each other. The bottom panels show the induced
scoring rule $S(p,q)$ defined by
\cref{eq:scoring-rule} or more specifically $S(p,q)= q S(p, 1) + (1 -
q) S(p, 0)$ at two different values of $q = 0.05,0.15$. For aesthetic
purposes, the scoring rules in \Cref{fig:S-pq-plot} are normalized
such that $-S(p,q) = 1$ and $-S(p,1-q) = 2$.

\Cref{fig:S} shows that the scoring rules $S_{\alpha,\beta}$, when $-1 \le \alpha,\beta \le 0$, are highly
sensitive to small differences of small probabilities. For example, in
\Cref{fig:S-plot} the loss function $-S_{\alpha,\beta}(p,1)$ is unbounded above when $\alpha,\beta \in
\{-1,0\}$, hence a small change of $p$ near $0$ may have a big impact
on the score. In \Cref{fig:S-pq-plot}, the
averaged scoring rules $S_{\alpha,\beta}(p,q)$, when
$(\alpha,\beta) = (-1, -1)$ or $(-1, 0)$, are also unbounded near $p =
0$.  Due to this reason, \citet[Section
2.6]{selten1998axiomatic} argued that these scoring rules are
inappropriate for probability forecast problems.

On the contrary, the unboundedness is actually a desirable feature for
propensity score estimation, as the goal is to avoid extreme
probabilities. Consider the standard inverse probability weights (IPW)
\begin{equation} \label{eq:ipw-ate}
  \hat{w}_i =
  \begin{cases}
    \hat{p}_i^{-1} & \mathrm{if}~T_i = 1, \\
    (1 - \hat{p}_i)^{-1} & \mathrm{if}~T_i = 0, \\
  \end{cases}
\end{equation}
where $\hat{p}_i = p_{\hat{\theta}}(X_i)$ is the estimated
propensity score for the $i$-th data point. This corresponds to
$\alpha = \beta = -1$ in the Beta family and estimates ATE. Several previous articles
\citep[e.g.][]{Robins2000,Kang2007,Robins2007} have pointed out the
hazards of using large inverse probability weights. For example, if the true propensity score
is $p(X_i) = q = 0.05$ and it happens that $T_i = 1$, we would want $\hat{p}_i$
not too close to $0$ so $\hat{w}_i$ is not too large. Conversely, we also
want $\hat{p}_i$ not too close to $1$, so in the more likely event that $T_i = 0$
the weight $\hat{w}_i$ is not too large either. In an \emph{ad hoc}
attempt to mitigate this issue, \citet{lee2011weight} studied weight truncation (e.g.\ truncate the
largest 10\% weights). They found that the truncation can reduce the
standard error of the estimator $\hat{\tau}$ but also increases the bias.

The covariate balancing scoring rules provide a more systematic
approach to avoid large weights. For example, the scoring rule $S_{-1,-1}$ precisely
penalizes large inverse probability weights as $-S_{-1,-1}(p,q)$ is unbounded above when
$p$ is near $0$ or $1$ (see the left plot in
\Cref{fig:S-pq-plot}). Similarly, when estimating the ATUT
$\tau_{-1,0}$, the weighting scheme would put $\hat{w}_i \propto (1 - \hat{p}_i) /
\hat{p}_i$ if $T_i = 1$ and $\hat{w}_i \propto 1$ if $T_i = 0$. Therefore we
would like $\hat{p}_i$ to be not close to $0$, but it is acceptable
if $\hat{p}_i$ is close to $1$. As shown in in
\Cref{fig:S-pq-plot}, the curve $-S_{-1,0}(p,q) = q / p +
(1 - q) \log (p / (1-p))$ precisely encourages this behavior, as it is unbounded
above when $p$ is near $0$ and grows slowly when $p$ is near $1$.



\bibliographystyle{chicago}
\bibliography{../reference/ref}

\end{document}